\documentclass[aps,twocolumn,PRL,reprint]{revtex4-1}

\usepackage{amsmath}
\usepackage{epsfig}
\usepackage{graphicx}
\usepackage{graphicx, color, epstopdf}
\usepackage{bm}
\usepackage{amssymb}
\usepackage{hyperref}
\usepackage{xcolor}
\usepackage{subfigure}
\usepackage{makecell}
\usepackage{tikz}

\usepackage{etoolbox}
\preto\section{%
  \ifnum\value{section}=4 %
  \fi
}

\hypersetup{hidelinks,
	colorlinks=true,
	allcolors=black,
	pdfstartview=Fit,
	breaklinks=true}
	
\definecolor{lime}{HTML}{A6CE39}
\DeclareRobustCommand{\orcidicon}{
\begin{tikzpicture}
\draw[lime, fill=lime] (0,0)
circle[radius=0.16]
node[white]{{\fontfamily{qag}\selectfont \tiny \.{I}D}};
\end{tikzpicture}
\hspace{-2mm}
}
\foreach \x in {A, ..., Z}{%
\expandafter\xdef\csname orcid\x\endcsname{\noexpand\href{https://orcid.org/\csname orcidauthor\x\endcsname}{\noexpand\orcidicon}}
}

\begin{document}
\title{{Quantum-Classical Correspondence of Non-Hermitian Symmetry Breaking}}
\author{Zhuo-Ting Cai}
\affiliation{National Laboratory of Solid State Microstructures, School of Physics,
Jiangsu Physical Science Research Center, and Collaborative Innovation Center of Advanced Microstructures,
Nanjing University, Nanjing 210093, China}
\author{Hai-Dong Li}
\affiliation{National Laboratory of Solid State Microstructures, School of Physics,
Jiangsu Physical Science Research Center, and Collaborative Innovation Center of Advanced Microstructures,
Nanjing University, Nanjing 210093, China}
\author{Wei Chen \hspace{-1.5mm}\orcidA{}}
\email{Corresponding author: pchenweis@gmail.com}
\affiliation{National Laboratory of Solid State Microstructures, School of Physics,
Jiangsu Physical Science Research Center, and Collaborative Innovation Center of Advanced Microstructures,
Nanjing University, Nanjing 210093, China}
\begin{abstract}
Real-to-complex spectral transitions
and the associated spontaneous symmetry breaking of eigenstates
are central to non-Hermitian physics, yet a comprehensive and universal theory that precisely describes the underlying physical mechanisms for each individual state remains elusive. Here, we resolve the mystery by employing the complex path integral formalism and developing a
generalized Gutzwiller trace formula.
These methodologies enable us to establish a universal quantum-classical
correspondence that precisely links the real or complex nature of individual energy levels
to the symmetry properties of their corresponding semiclassical orbits. Specifically, in systems with a general
$\eta$-pseudo-Hermitian symmetry, real energy levels are quantized
along periodic orbits that preserve the corresponding classical $S_\eta$ symmetry.
In contrast, complex conjugate energy levels arise from semiclassical orbits that
individually break the $S_\eta$ symmetry but together form $S_\eta$-symmetric
pairs. This framework provides a unified explanation for the spectral behaviors in
various continuous non-Hermitian models and for the $\mathcal{PT}$ transition in two-level systems.
Besides, we demonstrate that the exceptional point is inherently a quantum phenomenon, as it cannot be described by a single classical orbit.
Our work uncovers the physical mechanism of non-Hermitian symmetry breaking and introduces a new perspective with
broad implications for the control and application of non-Hermitian phenomena.
\end{abstract}

\date{\today}

\maketitle

\emph{Introduction.---}In the past decade, great research progress has been made in non-Hermitian
physics~\cite{bender2007making,Ashida2020NHphysic,El2018np}.
Different from the standard quantum mechanism in which the Hamiltonian of a closed system
is represented by a Hermitian operator, non-Hermitian Hamiltonians can effectively describe
various physical effects in open systems.
Recently, numerous interesting non-Hermitian phenomena have been discovered, such as the non-Hermitian spectral phase
transitions~\cite{Bender98prl,Guo2009prl,Ruter2010np,Regensburger2012Nature,Peng2014NP,LiangFeng2014science,ozdemir2019parity,feng2013nm,feng2017non,LonghiS2019prl1,
LonghiS2019prl2,LeeTonyE2014prl,XuePeng2022prl,chenshu2019prb,Jazaeri2001PRE,chenshu2021prb},
novel physics associated with exceptional points~\cite{Miri2019science,Dembowski2004pre,
Bergholtz2021rmp,chen2017exceptional}, the non-Hermitian skin effect~\cite{Wangzhong2018prl,
Kunst2018prl,Yokomizo2019prl,Borgnia20prl,Okuma2020prl,Zhangkai2020prl} and the non-reciprocal phase transitions~\cite{fruchart21nat},
which not only deepen our understanding of the physical world but may also lead to
interesting applications~\cite{Wiersig2014prl,Brandenbourger2019NC,Dong2019NE,WangZhong2021PRB,wangqiang2022Amplification,shao24prl,geng23prb}. These non-Hermitian effects
are directly attributable to, or closely linked with, the properties of
the non-Hermitian energy spectra and the corresponding eigenstates,
which constitute the main focus of non-Hermitian physics.

In contrast to the Hermitian Hamiltonian that ensures an entirely real energy spectrum,
non-Hermitian systems exhibit richer spectral characteristics. For the most relevant class of non-Hermitian Hamiltonians
that satisfy the condition of pseudo-Hermiticity~\cite{mostafazadeh2002pseudo,Ashida2020NHphysic},
their energy levels are either real or appear as complex conjugate pairs.
A notable example is the parity-time ($\mathcal{PT}$) transition~\cite{Bender98prl},
where variation in model parameters can drive a real-to-complex spectral transition,
accompanied by spontaneous symmetry breaking of the corresponding eigenstates.
These non-Hermitian phenomena have been known for a long time with
their validity grounded in mathematical theorems~\cite{wigner1960normal} and
confirmed by experiments~\cite{Guo2009prl,Ruter2010np,Regensburger2012Nature,Peng2014NP,LiangFeng2014science,ozdemir2019parity}.
However, unlike the spontaneous symmetry breaking
in conventional phase transitions stabilized by the principle of lowest free energy,
the non-Hermitian spectral transitions {and symmetry breaking}
just take place naturally as the model parameters vary.
Whether a fundamental and universal mechanism exists governing these phenomena at the level of individual states in both symmetric and symmetry-breaking phases remains an open question~{\cite{difference}}.
Answering this question can not only enhance our comprehension of non-Hermitian phenomena
but also provide practical insights for their {precise} control.

\begin{figure}[!htb]
\centering
\includegraphics[width=\columnwidth]{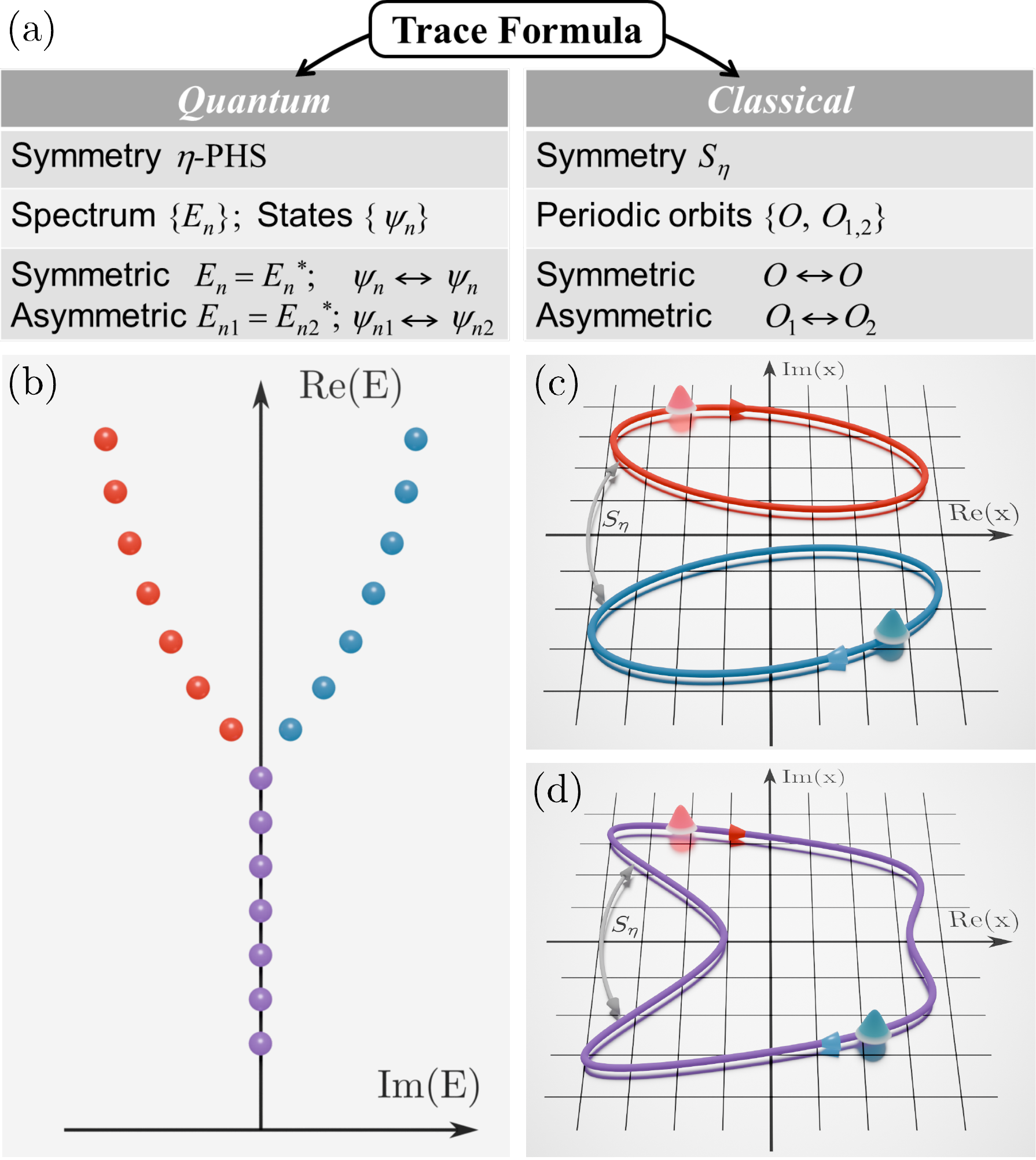}
\caption{Quantum-semiclassical correspondence of
non-Hermitian physics. (a) Quantum and classical properties are related by the trace formula.
(b)-(d) Schematic diagrams illustrating the correspondence between non-Hermitian energy spectrum
and semiclassical orbits of wave packets.
(b) Energy spectrum consisting of real parts and complex conjugate pairs represented by colored balls.
(c) Two semiclassical orbits in complex coordinate space forming an $S_\eta$-symmetric pair,
corresponding to the complex energy pairs.
(d) An $S_\eta$-symmetric orbit corresponding to real eigenenergies.
The eigenenergies and the corresponding orbits are indicated by the same colors.
The $S_\eta$ symmetry operation is denoted by the gray bidirectional arrows.
}
\label{fig1}
\end{figure}

{In this Letter, we provide an affirmative answer to this question by developing a general theory
of non-Hermitian symmetry breaking. In particular, we establish a quantum-classical correspondence
between the properties of individual energy levels and
the symmetry of their corresponding semiclassical orbits}, as illustrated in Fig.~\ref{fig1}.
This correspondence is achieved by employing the complex path integral approach and
deriving a generalized Gutzwiller trace formula
for a broad class of analytic non-Hermitian problems exhibiting pseudo-Hermiticity. We demonstrate that
any $\eta$-pseudo-Hermitian symmetry ($\eta$-PHS) (Eq.~\eqref{hs}, defined for the quantum Hamiltonian)
leads to a corresponding semiclassical $S_\eta$ symmetry (Eq.~\eqref{sym_Hcl}, defined
for the corresponding classical Hamiltonian).
The trace formula quantizes each energy level
along periodic orbits in complex spacetime, and the symmetry properties of the orbits
dictate whether the energy level is real or complex.
Specifically, a real energy level $E_n$ occurs as the semiclassical orbit $O$ possesses the $S_\eta$ symmetry.
Otherwise, complex conjugate energy pairs, denoted as $E_{n1}=E^*_{n2}$, arise
when the corresponding orbits $O_{1}$ and $O_{2}$ individually break the $S_\eta$ symmetry
but collectively form an $S_\eta$-symmetric pair [cf. Fig~\ref{fig1}].
Besides the Quantum-Classical correspondence, we also demonstrate that the exceptional point is inherently a quantum phenomenon, which cannot be described by a single classical orbit.
The validity and universality of our theory are confirmed through its
application to various continuous non-Hermitian models and the $\mathcal{PT}$ transition in two-level systems.
Our work offers a thorough and comprehensive understanding of the non-Hermitian spectrum, applicable to a wide range of non-Hermitian systems and general pseudo-Hermitian symmetries.

\emph{Quantum-classical correspondence of symmetry.---}
We first investigate a general
non-Hermitian Hamiltonian $\mathcal{H}(\hat{x}, \hat{p})$ with a continuous energy spectrum,
where $\hat{x}$ and $\hat{p}$ denote the coordinate and momentum operators, respectively.
Non-Hermitian Hamiltonians that satisfy the condition of
$\eta$-pseudo-Hermiticity~\cite{mostafazadeh2002pseudo} are of particular interest:
\begin{equation}\label{hs}
\eta \mathcal{H}(\hat{x}, \hat{p}) \eta^{-1}=\mathcal{H}^{\dagger}(\hat{x}, \hat{p}),
\end{equation}
where $\eta = \eta^\dagger$ is a general invertible Hermitian operator.
This constraint, referred to as $\eta$-PHS,
plays a critical role in shaping the energy spectrum, with its preservation
or breaking strongly influencing the spectral properties~\cite{wigner1960normal}.

To establish the connection between quantum and
semiclassical pictures in Fig.~\ref{fig1}, we start with the propagator
$G\left(x_f, x_i, t\right)=\left\langle x_{f}|e^{-i \mathcal{H} t}| x_{i}\right\rangle$ from
$x_i$ to $x_f$ in coordinate space ($\hbar=1$ is taken).
For later purposes, all physical quantities, including coordinates,
momenta, and time, are considered to be complex~\cite{mclaughlin1972complex,ShaoKai2022prb,yang24prb}.
It means that $G\left(x_f, x_i, t\right)$ is expressed under the biorthonormal eigenbasis of $\hat{x}$.
The $\eta$-PHS~\eqref{hs} imposes the following constraint on the propagator~\cite{sm}
\begin{equation}\label{gg}
G_\eta\left(x_f, x_i, t\right)=G^*(x_i^*, x_f^*,-t^*),
\end{equation}
where $G_\eta\left(x_f, x_i, t\right)=\langle x_{f}|e^{-i \mathcal{H}\left(\hat{x}_\eta, \hat{p}_\eta\right) t}| x_{i}\rangle$ with
$\hat{x}_\eta=\eta\hat{x}\eta^{-1}, \hat{p}_\eta=\eta\hat{p}\eta^{-1}$,
and $G^*(x_i^*, x_f^*,-t^*)=\left[\langle x_{i}|e^{i \mathcal{H}(\hat{x}^{\dagger}, \hat{p}^{\dagger}) t^*}| x_{f}\rangle\right]^*$.
Such a constraint indicates that any specific $\eta$-PHS
corresponds to a classical symmetry $S_\eta$.
To this end, we express the propagator in terms of the path integral in phase space
and Eq.~\eqref{gg} leads to
\begin{equation}\label{path}
     \int_{x_i}^{x_f}\mathcal{D}[\xi] e^{i \int_0^t\left[p \dot{x}-H\left(x_\eta, p_\eta\right)\right] d t^{\prime}}
     =\int_{x_i}^{x_f}\mathcal{D}[\xi] e^{i \int_0^t\left[p \dot{x}-H^*\left(x^*, p^*\right)\right] d t^{\prime}},
\end{equation}
where $\xi(t):=\{x(t),p(t)\}$ denotes the phase space coordinate,
the integrations over $x$, $p$ and $t'$ are along certain paths in the respective complex planes,
and $x_\eta=x_\eta(x,p), p_\eta=p_\eta(x,p)$ are both functions of $x,p$~\cite{sm}.
In the path integral formula, the coordinates, momenta and thus the Hamiltonian $H$ all become c numbers.
Equation.~\eqref{path} holds for arbitrary values of $x_i,x_f$ and
$t$, indicating that the semiclassical Hamiltonian
must adhere to the following constraint or, referred to as the $S_\eta$ symmetry
\begin{equation}\label{sym_Hcl}
   S_\eta: H(x_\eta,p_\eta)=H^*(x^*,p^*),
\end{equation}
which is the semiclassical counterpart of the $\eta$-PHS.
The $S_\eta$ symmetry governs the semiclassical dynamics of the system
and will be inherited by the classical orbits.

\emph{Generalized Gutzwiller trace formula.---}The energy spectrum of $\mathcal{H}$ corresponds to the poles of
the trace of the Green's operator $G(E)=\text{Tr}(\mathcal{H}-E)^{-1}$.
This conclusion applies to both real and complex energy values,
indicating that the energy spectrum of both Hermitian and non-Hermitian systems can be analyzed in the same way.
In the Hermitian regime, the Gutzwiller trace formula expresses $G(E)$ in terms of the contour integral
over periodic orbits so as to establish a connection between
the energy spectrum and these semiclassical orbits~\cite{gutzwiller1971JMP}.
We demonstrate that, by generalizing the Gutzwiller trace formula to the non-Hermitian regime,
the long-standing puzzle of the underlying scenario for
non-Hermitian spectral properties and symmetry breaking of eigenstates
can be resolved through the semiclassical perspective.

Interpreting $G(E)$ as the Fourier transform of the time-domain propagator
with the trace conducted in the coordinate space yields the complex path integral formula
\begin{equation}
G(E)=i\int dt \ e^{i E t}\int dx_i \int_{x_i}^{x_i} \mathcal{D}[\xi] e^{i\int_0^t[p\dot{x}-H(x,p)]dt'}.
\end{equation}
We extend the conventional Gutzwiller trace formula to the non-Hermitian regime by repeatedly
applying saddle point approximation (SPA)~\cite{bender2013advanced} to
the complex path integral, instead of the stationary phase approximation in the Hermitian regime~\cite{sm}.
First, applying SPA to $\int_{x_i}^{x_i}\mathcal{D}[\xi]$
reduces the path integral to the contributions by those
closed orbits obeying the complex Hamilton's canonical equations
$\dot{x}={\partial H}/{\partial p}, \dot{p}=-{\partial H}/{\partial x}$.
Second, SPA applied to $\int dx_i$ selects out the periodic orbits
with identical initial and final momenta
at $x_i$, including both their real and imaginary parts.
Finally, SPA applied to $\int dt$ establishes the relation function $E=E(T)$ between
energy $E$ and periodicity $T$ of the classical motion.
In the non-Hermitian regime, periodic orbits may not exist
when the periodicity $T$ is restricted to be real~\cite{ShaoKai2022prb}. To obtain periodic orbits, a complex $T$
should be generally assumed~\cite{yang24prb}.
The time-evolution contour lies in the complex $t$-plane that connects 0 and $T$.
Although there are an infinite number of such contours,
the evaluation of $G(x_i,x_i,T)$ does not rely on
the specific choice of the contour~\cite{mclaughlin1972complex}. Therefore,
by choosing a proper time contour, $G(E)$ and the spectrum can be determined without ambiguity.
After three steps of SPAs, we arrive at the trace formula
involving contour integrals along all periodic orbits $O$ as~\cite{sm}
\begin{equation}\label{G_semi}
    G(E)=i T(E)\frac{e^{i\left(\oint_{O} p d x- 2\pi\mu\right)}}
    {1-e^{i\left(\oint_{O} p d x-2\pi \mu\right)}},
\end{equation}
where $T(E)$ is the periodicity of the
orbit for a given energy $E$ and
$-4\mu$ defines the Maslov index~\cite{rajaraman1982solitons}. Extending the formula to more general cases involving
distinct categories of periodic orbits is straightforward~\cite{sm}.
Although Eq.~\eqref{G_semi} possesses the same form as that in the Hermitian regime~\cite{gutzwiller1971JMP}, the key distinction is
that all physical quantities now reside in the complex domain.
Poles of Eq.~\eqref{G_semi} correspond to the quantization condition in complex spacetime as
\begin{equation}\label{QC}
    \oint_{O} p dx=(n+\mu)2\pi,
\end{equation}
which determines the eigenenergies of the system.
From Eq.~\eqref{G_semi}, one can understand that
the imaginary part of $\oint_{O} p dx$ represents the growth or decay of wave
over one propagating cycle, while its real part represents the accumulated phase.
Therefore, the quantization condition~\eqref{QC} determines the eigenenergies
by selecting periodic orbits that exhibit
quantized phase accumulations while maintaining a constant amplitude.

\emph{Semiclassical interpretation of non-Hermitian spectrum.---}We have established the quantum-classical correspondence
of symmetries ($\eta$-PHS~$\leftrightarrow$~$S_\eta$ as shown in Fig.~\ref{fig1}) and interpreted
non-Hermitian energy levels by quantized semiclassical orbits.
Now we prove that the real or complex nature of
each energy level is precisely determined by the $S_\eta$-symmetric properties of the corresponding periodic orbits.
Physically, the semiclassical symmetry $S_\eta$ in Eq.~\eqref{sym_Hcl} implies the following~\cite{sm}
\par (i)
If $\xi_1(t)=\xi(t)$ is a solution of the canonical equation,
$\xi_2(t^*)=\xi_\eta[\xi^*(\mp t)]$ is also a solution but along
the time contour $t^*$, the two being $S_\eta$-symmetric to each other,
where the ``$\mp$'' corresponds to whether the $\eta$ operation involves a transpose or not.

Moreover, for periodic orbits $O_1(t)$ and $O_2(t^*)$ forming $S_\eta$-symmetric pairs,
the following conclusions can be proved:
\par(ii) Their contour integrals satisfy the equality
\begin{equation}
    \oint_{O_2(t^*)} p_2(t^*) dx_2(t^*)=\left(\oint_{O_1(t)} p_1(t) dx_1(t)\right)^*,
\end{equation}

\par(iii) The energies and periods of $O_1$ and $O_2$ are complex conjugates:
\begin{equation}
    E_{1}=E^*_{2},\ \  T_{1}(E_1)=T^*_{2}(E_2).
\end{equation}

The combination of these properties and the quantization condition
can only yield two possible outcomes for nondegenerate
eigenenergies~\cite{sm}:
\par(1) $O_1$ and $O_2$ are essentially the same orbit obeying the $S_\eta$ symmetry,
and the quantized eigenenergies satisfying $E_n=E_n^*$, are real;
\par(2) $O_1$ and $O_2$ are different orbits related by the $S_\eta$ operation ($S_\eta: O_1\leftrightarrow O_2$),
and the quantized
eigenenergies constitute complex conjugate pairs with $E_{n1}=E_{n2}^*$.

\begin{figure*}[!htb]
    \centering
        \includegraphics[width=1\textwidth]{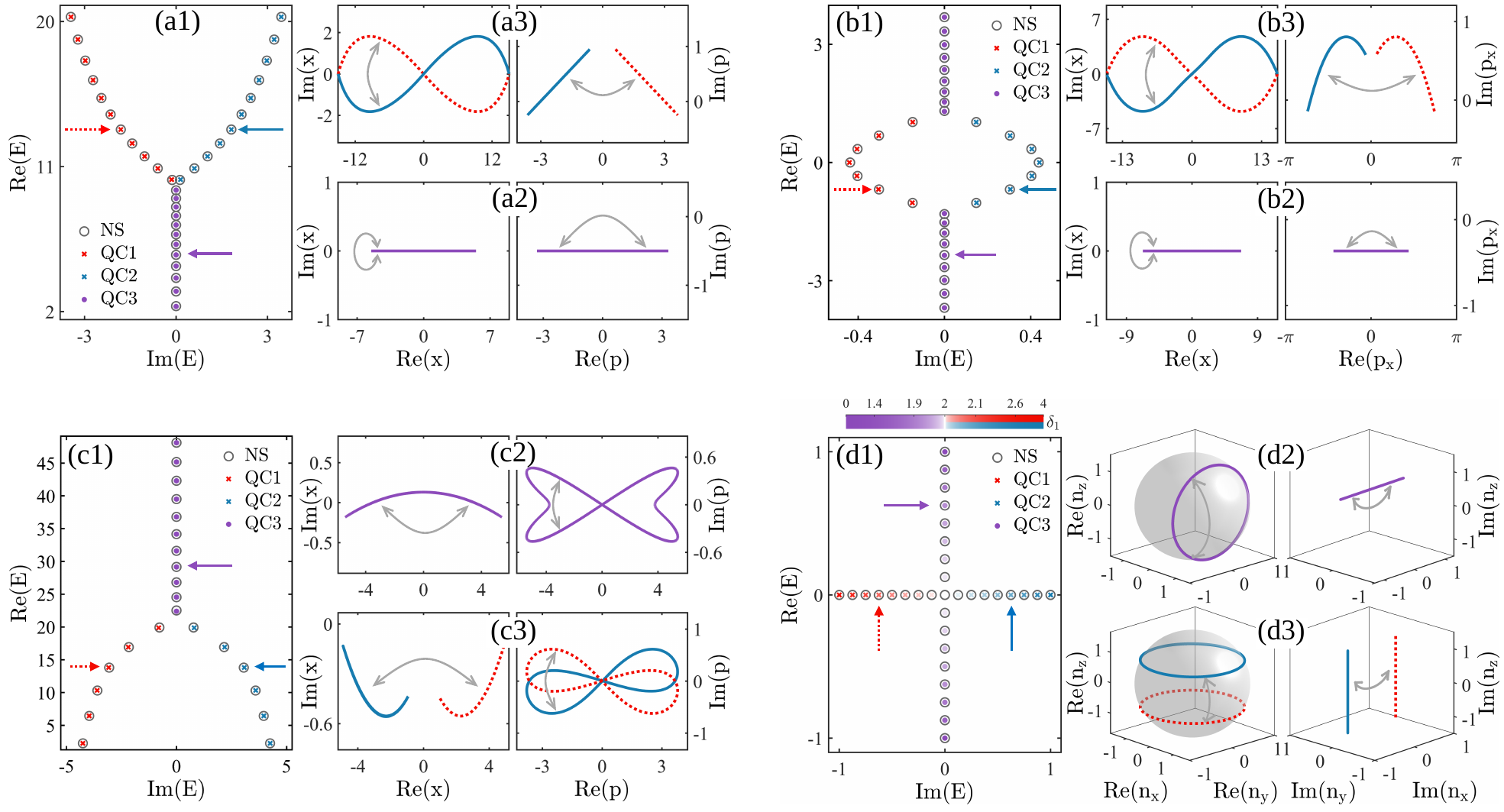}
        \caption{Quantum-semiclassical correspondence validated by four concrete examples.
        (a1) Comparison between energy spectra obtained by
        numerical solutions (NS) of the eigenvalue equations of
        $\mathcal{H}_1$ and those solved by the quantization conditions (QC1,QC2,QC3) applied to
        different periodic orbits in (a2, a3). (a2) $S_\eta$-symmetric orbits
        in phase space corresponding to real eigenenergies in (a1). (a3) Asymmetric orbits in phase space constitute an $S_\eta$-symmetric pair
        corresponding to complex eigenenergies in (a1). The gray bidirectional arrows
        denote specific $S_\eta$ symmetry operations. The results for $\mathcal{H}_2, \mathcal{H}_3$ and $\mathcal{H}_4$
        are presented in the parallel way in (b1)-(b3), (c1)-(c3) and (d1)-(d3), respectively.
        (d1) Energy spectrum of the two-level system evolves with varying $\delta_1$.
        The relevant parameters are:
        (a1)-(a3) $\gamma=0.5,V_0=1,L=30$, (b1)-(b3) $t_0=-1,\delta=0.35,L=32$, (c1)-(c3) $a=3.5,g=0.1,\Gamma=1.25$, (d1)-(d3) $t_1=2$,
        where $L$ denotes the scales of the systems.
        }
        \label{fig2}
    \end{figure*}

Through this method, the non-Hermitian spectrum can be well interpreted
by assigning a classical perspective to
each energy level as sketched in Fig.~\ref{fig1}.
In the following, we use several concrete examples to verify the general conclusions
obtained above and show the results collectively in Fig.~\ref{fig2}.
The main information conveyed by Fig.~\ref{fig2} is the
quantum-classical correspondence between the energy spectrum properties and the symmetries of the corresponding orbits
(more calculation details are provided in the Supplemental Material~\cite{sm}).

\emph{Continuous systems.---}
First, we consider a continuous model
$
\mathcal{H}_1=(\hat{p}+i\gamma)^2+V(\hat{x})
$
that exhibits non-Hermitian skin effect~\cite{Wangzhong2018prl,
Kunst2018prl,Yokomizo2019prl,Borgnia20prl,Zhangkai2020prl,Okuma2020prl},
where the real $\gamma$ measures the strength of the skin effect and $V(\hat{x})=V_0|\hat{x}|$ is a Hermitian potential.
It possesses time-reversal ($\mathcal{T}$) symmetry with $\mathcal{H}_1=\mathcal{T}\mathcal{H}_1\mathcal{T}^{-1}
=\mathcal{H}_1^*$, which is equivalent to the $\mathsf{T}$-PHS:
$\mathsf{T}\mathcal{H}_1\mathsf{T}^{-1}=\mathcal{H}_1^{\dagger}$, with
$\mathsf{T}$ the transpose operation.
Periodic boundary conditions are adopted in the calculations.
We compare in Fig.~\ref{fig2}(a1) the spectrum obtained by
numerically solving the eigenvalue equation for $\mathcal{H}_1$
(with real $x$ and $\hat{p}=-i\partial_x$) and that acquired
via the generalized Gutzwiller trace formula using the quantization condition~\eqref{QC}. The results from both
methods demonstrate remarkable agreement, affirming the validity of our theory.
The energy spectrum contains two distinctive parts,
the one with real values and the remaining with complex conjugate pairs.
Whether the eigenvalue is real or complex depends on the symmetry of the corresponding orbit.
The $S_\eta$ symmetry~\eqref{sym_Hcl} of the semiclassical Hamiltonian
is specified as $S_{\mathsf{T}}: H_1(x,-p)=H_1^*(x^*,p^*)$ in this case. This condition ensures that the solutions for periodic orbits,
$O_1(t)=\{x(t), p(t)\}$ and $O_2(t^*)=\{x^*(-t), -p^*(-t)\}$, always form $S_{\mathsf{T}}$-symmetric pairs.
For every real energy level, the corresponding orbit obeys the $S_{\mathsf{T}}$ symmetry, as shown in Fig.~\ref{fig2}(a2).
In contrast, for any complex conjugate pair of energy levels, each of the corresponding
orbits in phase space breaks the $S_{\mathsf{T}}$ symmetry. However,
they are related to each other through the symmetry operation as $S_\mathsf{T}: O_1\leftrightarrow O_2$, as illustrated by
the gray bidirectional arrows in Fig.~\ref{fig2}(a3).
The results can be understood as follows.
A particle with low energies is confined within the potential well
with its orbit exhibiting the $S_{\mathsf{T}}$ symmetry, resembling the situation for
an open boundary condition. At high energies, the particle can overcome and traverse
the potential barrier, which drastically changes the periodic orbit and breaks its symmetry.
Two additional examples are investigated in parallel, with the results
depicted in Figs.~\ref{fig2}(b1-b3) and \ref{fig2}(c1-c3) respectively,
both leading to the same conclusion. Detailed discussions can be found in the Appendix.

\emph{$\mathcal{PT}$ transition in two-level systems.---}
Apart from continuous models, the non-Hermitian spectrum
for discrete systems can also be interpreted within our framework.
Take the $\mathcal{PT}$ transition in the non-Hermitian two-level systems as an example,
which has been extensively explored in optical systems with gain and loss~\cite{Guo2009prl,Ruter2010np,
Regensburger2012Nature,Peng2014NP,LiangFeng2014science,ozdemir2019parity}.
It can be effectively captured by
\begin{equation}\label{spin}
\mathcal{H}_4=\frac{1}{2}\bm{M}\cdot \bm{\sigma},\ \ \bm{M}=(t_1,0,i\delta_1),\ \ \bm{\sigma}=(\sigma_x,\sigma_y,\sigma_z),
\end{equation}
where $t_1$ is the hopping, $i\delta_1$ describes the gain and loss for each level
and $\bm{\sigma}$ is a vector composed of three Pauli matrices $\sigma_{x,y,z}$.
In optics, the two levels correspond to two different sites, which are interchanged under the parity operation
$\mathcal{P}=\sigma_x$. Meanwhile, time reversal operation $\mathcal{T}$ inverts the gain and loss.
The system only obeys the combined $\mathcal{PT}$ symmetry: $\mathcal{PT}\mathcal{H}_4(\mathcal{PT})^{-1}
=\sigma_x \mathcal{H}_4^*\sigma_x=\mathcal{H}_4$, or
the $\mathcal{P}\mathsf{T}$-PHS: $\mathcal{P}\mathsf{T}\mathcal{H}_4(\mathcal{P}\mathsf{T})^{-1}=\mathcal{H}_4^{\dagger}$.
The energy spectrum as a function of \(\delta_1\), shown in Fig.~\ref{fig2}(d1), remains real for \(\delta_1<t_1\)
but becomes complex when \(\delta_1>t_1\).
The spectral properties can be explained by the semiclassical (pseudo-)spin dynamics.
For this purpose, we derive a similar trace formula for two-level systems applying the technique of spin path integral~\cite{sm,Nielsen1988NPB,altland2010condensed}.
The corresponding semiclassical $S_{\mathcal{P}\mathsf{T}}$ symmetry is expressed as
\begin{equation}\label{sym_spin}
S_{\mathcal{P}\mathsf{T}}: H_4(n_x,n_y,-n_z)=H_4^*(n_x^*,n_y^*,n_z^*),
\end{equation}
where the vector $\bm{n}=(n_x, n_y, n_z)$
denotes the orientation of the classical spin.
It dominates the classical equation of motion $\dot{\bm{n}}=\bm{M}\times\bm{n}$
and ensures the existence of two solutions
$\bm{n}_1(t)=[n_x(t),n_y(t),n_z(t)]$ and $\bm{n}_2(t^*)=[n_x^*(-t),n_y^*(-t),-n_z^*(-t)]$
forming $S_{\mathcal{P}\mathsf{T}}$-symmetric pairs.
When $\delta_1<t_1$, the average spin vector during one period is aligned with $\pm \bm{M}$,
and the periodic trajectories traced by its tip exhibit
the $S_{\mathcal{P}\mathsf{T}}$ symmetry; see Fig.~\ref{fig2}(d2).
As $\delta_1$ crosses $t_1$, the classical trajectories of spin
diverge to infinity, accompanied by an abrupt change in the direction of the average spin vector
from $\pm \bm{M}$ to $\pm i\bm{M}$,
which signifies the critical point of the $\mathcal{PT}$ transition~\cite{sm}.
For $\delta_1>t_1$, the average spin vector remains aligned with $\pm i \bm{M}$,
resulting in $S_{\mathcal{P}\mathsf{T}}$-symmetric pairs of spin orbits,
where each orbit individually breaks the $S_{\mathcal{P}\mathsf{T}}$ symmetry; see Fig.~\ref{fig2}(d3).

For continuous models, the quantization conditions applied along individual semiclassical orbits
yield an accurate energy spectrum across most energy regions, except for small deviations near the transition point, known as the exceptional point.
These deviations can be corrected within the general theoretical framework of complex path integrals by accounting for the pronounced tunneling between orbits~\cite{sm}. For the two-level system, the quantization conditions accurately determine the energy spectrum near the $\mathcal{PT}$ transition point. However, at this point, the classical orbits diverge and change abruptly. These results indicate that the exceptional point is inherently a quantum phenomenon, as it cannot be described by a single classical orbit~\cite{sm}.

\emph{Summary and outlook.---}
{We have uncovered the general physical mechanism governing non-Hermitian spectral properties and symmetry breaking.}
The intuitive physical picture presented here not only enhances our comprehension of
non-Hermitian phenomena but also offers practical insights for their manipulation.
This framework operates within the general context of non-Hermitian physics,
making it applicable to a wide range of areas in physics, such as quantum physics,
condensed matter physics, optics and acoustics.
Demonstrating the quantum-classical correspondence in various physical systems shows great
promise by comparing quantum eigenenergies with the corresponding semiclassical dynamics
across different parametric regions. Meanwhile, extending the current theory
from single-particle to many-body scenarios is also of significant interest.

\begin{acknowledgments}
    We thank Gu Zhang, Weiyin Deng,
 and Feng Mei for the helpful feedback on our manuscript.
    This work was supported by
    the National Natural Science Foundation of
    China (No.  12222406), the Natural Science Foundation of Jiangsu Province (No. BK20233001),
    the Fundamental Research Funds for the Central Universities (No. 2024300415), and
    the National Key Projects for Research and Development of China (No. 2022YFA1204701).
\end{acknowledgments}

%


\hspace*{\fill}

\ \ \ \ \ \ \ \ \ \ \ \ \ \ \ \ \ \ \ \ \ \ \ \textbf{End Matter}

\emph{Appendix.---}The second example is the nonreciprocal lattices subject to a magnetic field~\cite{ShaoKai2022prb}, described
by the Hamiltonian
$
\mathcal{H}_2=-2[t_0\cos \hat{p}_x + i\delta_x \sin \hat{p}_x + t_0 \cos(p_y-B\hat{x})],
$
where Fourier transform is performed in the $y$ direction with $p_y$ the corresponding momentum
and $\delta_x$ measures the nonreciprocity of hopping in the $x$ direction.
In this way, the Hamiltonian reduces to
an effective 1D problem with non-Hermitian skin effect. The magnetic field
manifests as a potential modulation, with the dynamics of particle
resembling the 1D projection of the cyclotron motion in 2D space.
The real energy levels in the long-wavelength limit were explained in Ref.~\cite{ShaoKai2022prb},
yet an interpretation of the remaining part is still lacking.
The 2D system possesses the combined mirror-time reversal symmetry,
which can be interpreted as the $ \mathcal{M}\mathsf{T}$-PHS: $\mathcal{M}\mathsf{T}
\mathcal{H}_2(\mathcal{M}\mathsf{T})^{-1}=\mathcal{H}_2^{\dagger}$,
where the mirror reflection $\mathcal{M}$ is about the $x$ axis.
The spectral properties are dictated by the $\mathcal{M}\mathsf{T}$-PHS and its classical counterpart $S_{\mathcal{M}\mathsf{T}}$.
To verify this, we adopt the periodic boundary condition in the $x$ direction by choosing a proper magnetic field
and compare the energy spectra obtained by two approaches in Fig.~\ref{fig2}(b1).
The semiclassical 1D Hamiltonian obeys the symmetry
$S_{\mathcal{M}\mathsf{T}}: H_2\left(x-p_y/B,-p_x\right)=H_2^*\left(x^*-p_y/B,p_x^*\right)$, where $ p_y/B $ is real.
It reduces to $H_2(x,-p_x) = H_2^*(x^*,p_x^*)$ through the replacement $ x \rightarrow x + {p_y}/{B} $.
Accordingly, two solutions of $S_{\mathcal{M}\mathsf{T}}$-symmetric orbits $O_1(t)=\{x(t), p_x(t)\}$ and $O_2(t^*)=\{x^*(-t), -p_x^*(-t)\}$
come up in pairs. In particular,
the motion of a particle (hole) near the band bottom (top) exhibits $S_{\mathcal{M}\mathsf{T}}$ symmetry,
giving rise to real Landau levels after quantization; see Fig.~\ref{fig2}(b2);
However, for energies away from band edges the particle can traverse the potential, resulting in $S_{\mathcal{M}\mathsf{T}}$-symmetric
pairs of semiclassical orbits as well as complex eigenvalues, as shown in Fig.~\ref{fig2}(b3).

For the third example, we consider a particle moving within a double-well potential
that is subject to an antisymmetric gain and loss effect described by the Hamiltonian
$
\mathcal{H}_3=\hat{p}^2+g(\hat{x}^2-a^2)^2+i\Gamma \hat{x},
$
where $g,a,\Gamma$ are real parameters. The model possesses
$\mathcal{PT}$ symmetry, or equivalently, the
$\mathcal{P}\mathsf{T}$-PHS: $\mathcal{P}\mathsf{T}\mathcal{H}_3(\mathcal{P}\mathsf{T})^{-1}=\mathcal{H}_3^{\dagger}$.
The energy spectrum is shown in Fig.~\ref{fig2}(c1),
characterized by high-energy real eigenvalues and low-energy complex conjugate pairs.
Its semiclassical interpretation is reflected by the symmetry
$S_{\mathcal{P}\mathsf{T}}: H_3(-x,p)=H_3^*(x^*,p^*)$.
Such a constraint by symmetry dictates the equation of motion
and leads to two series of solutions $O_1(t)=\{x(t), p(t)\}$ and $O_2(t^*)=\{-x^*(-t), p^*(-t)\}$,
which form $S_{\mathcal{P}\mathsf{T}}$-symmetric pairs.
At high energies, particles move freely in the whole region and the gain and loss effects
compensate each other. Therefore, the orbits possess the $S_{\mathcal{P}\mathsf{T}}$ symmetry
and the eigenenergies are real after quantization; see Fig.~\ref{fig2}(c2).
At low energies, however, particles are trapped in the left or right potential wells,
experiencing unbalanced gain and loss. As a result, each semiclassical
orbit breaks the $S_{\mathcal{P}\mathsf{T}}$ symmetry individually, but
together they form $S_{\mathcal{P}\mathsf{T}}$-symmetric pairs, as illustrated in Fig.~\ref{fig2}(c3).

\end{document}


\title{Supplemental Material for ``Quantum-Classical Correspondence of Non-Hermitian Symmetry Breaking''}

\author{Zhuo-Ting Cai}
\affiliation{National Laboratory of Solid State Microstructures, School of Physics,
Jiangsu Physical Science Research Center, and Collaborative Innovation Center of Advanced Microstructures,
Nanjing University, Nanjing 210093, China}
\author{Hai-Dong Li}
\affiliation{National Laboratory of Solid State Microstructures, School of Physics,
Jiangsu Physical Science Research Center, and Collaborative Innovation Center of Advanced Microstructures,
Nanjing University, Nanjing 210093, China}
\author{Wei Chen \hspace{-1.5mm}\orcidA{}}
\email{Corresponding author: pchenweis@gmail.com}
\affiliation{National Laboratory of Solid State Microstructures, School of Physics,
Jiangsu Physical Science Research Center, and Collaborative Innovation Center of Advanced Microstructures,
Nanjing University, Nanjing 210093, China}

\maketitle

\onecolumngrid
\renewcommand{\theequation}{S.\arabic{equation}}
\setcounter{equation}{0}
\renewcommand{\thefigure}{S.\arabic{figure}}
\setcounter{figure}{0}

\section{restriction on the propagator by $\eta$-pseudo-Hermiticity}
The $\eta$-pseudo-Hermitian symmetry ($\eta$-PHS) of the Hamiltonian [Eq.~(1) of the main text]
\begin{equation}\label{etasym}
\eta \mathcal{H} \eta^{-1}=\mathcal{H}^{\dagger},
\end{equation}
imposes the following restriction on the propagator [Eq.~(2) of the main text]
\begin{equation}\label{g}
G_\eta\left(x_f, x_i, t\right)=G^*(x_i^*, x_f^*,-t^*),
\end{equation}
with $G_\eta\left(x_f, x_i, t\right)=\langle x_{f}|e^{-i \mathcal{H}\left(\hat{x}_\eta, \hat{p}_\eta\right) t}| x_{i}\rangle$
and $G^*(x_i^*, x_f^*,-t^*)=\left[\langle x_{i}|e^{i \mathcal{H}(\hat{x}^{\dagger}, \hat{p}^{\dagger}) t^*}| x_{f}\rangle\right]^*$.
This can be obtained as follows
\begin{equation}
\begin{split}
G_\eta\left(x_f, x_i, t\right)&=\langle x_{f}|e^{-i \mathcal{H}\left(\hat{x}_\eta, \hat{p}_\eta\right) t}| x_{i}\rangle
=\langle x_{f}|e^{-i \mathcal{H}^\dag\left(\hat{x}, \hat{p}\right) t}| x_{i}\rangle
=\left[\langle x_{i}|\left(e^{-i \mathcal{H}^\dag\left(\hat{x}, \hat{p}\right) t}\right)^\dag| x_{f}\rangle\right]^*\\
&=\left[\langle x_{i}|e^{i \mathcal{H}\left(\hat{x}^\dag, \hat{p}^\dag\right) t^*}| x_{f}\rangle\right]^*=G^*(x_i^*, x_f^*,-t^*),
\end{split}
\end{equation}
where we have used the properties of the biorthonormal eigenbases ($|x_i\rangle=|x_i\rangle_R, |x_f\rangle=|x_f\rangle_L$ with $L,R$ denoting
the left and right eigenvectors) for $\hat{x}$
\begin{equation}
\hat{x}|x_i\rangle=x_i|x_i\rangle,\ \  \langle x_i|\hat{x}^\dag=\langle x_i|x^*_i,\ \  \langle x_f| \hat{x}=\langle x_f| x_f,\ \
\hat{x}^\dag|x_f\rangle= x_f^*|x_f\rangle.
\end{equation}
It should be noted that in Eq.~\eqref{etasym}, the Hermitian conjugation
operating on $\mathcal{H}$ \emph{does not} act on individual variables $\hat{x}, \hat{p}$.

This relation can also be derived through analytical continuation.
In the realm of complex analysis, the behavior of analytic functions along the real
axis is sufficient to deduce their properties across the entire complex plane.
For real $x$, if two functions $f(x),g(x)$ fulfill the condition $f(x)=g^*(x)$,
then their $n$th-order derivatives at $x=0$, denoted as $f^{(n)}(0)$ and $g^{(n)}(0)$,
fulfill the condition $f^{(n)}(0)=\left[g^{(n)}(0)\right]^*$.
Consequently, for complex $x$, we have
\begin{equation}
f(x)=\sum_n \frac{f^{(n)}(0)}{n!} x^n=\sum_n \left[\frac{g^{(n)}(0)}{n!} (x^{*})^n\right]^*=g^*(x^*).
\end{equation}
We apply this property to our problem. Specifically, in real spacetime, the $\eta$-PHS of the Hamiltonian
leads to
\begin{equation}
G_\eta(x_f,x_i,t)=\langle x_{f}|e^{-i \mathcal{H}\left(\hat{x}_\eta, \hat{p}_\eta\right) t}| x_{i}\rangle=\left[\langle x_{i}|e^{i \mathcal{H}(\hat{x}, \hat{p}) t}| x_{f}\rangle\right]^*=G^*(x_i,x_f,-t).
\end{equation}
An analytical continuation of this equality to complex spacetime yields Eq.~\eqref{g}.

\section{complex path integral and classical $S_\eta$ symmetry}
We start from the propagator $G\left(x_f, x_i, t\right)=\langle x_f|e^{-i\mathcal{H}t}|x_i\rangle$ in
complex spacetime to derive the complex path integral formula.
When the phase space extends from real to complex domain,
biorthonormal eigenbases for both coordinates and momenta are generally considered (we drop the labels for the
left and right eigenvectors for brevity),
and the completeness relations involve double integrals as $\int dx_R dx_I |x \rangle \langle x|=1$ and $\int dp_R dp_I |p \rangle \langle p|=1$,
with both $x=x_R+ix_I$ and $p=p_R+ip_I$ being complex. Following standard path integral procedure
by decomposing the entire time evolution along certain time
contour in the complex plane into small intervals $\epsilon$ and inserting the completeness relations properly yields
\begin{equation}
\begin{aligned}
 G\left(x_f, x_i, t\right)=\int \prod_{n=1}^{N-1} dx_{R,n}dx_{I,n}\prod_{n=1}^{N} dp_{R,n}dp_{I,n}
\exp\left\{i\sum_{n=1}^{N}\left[p_n(x_{n}-x_{n-1})-\epsilon H(p_n,x_n)\right]\right\}.
 \end{aligned}
\end{equation}
It is convenient to use complex variables as
$x_n=x_{R,n}+ix_{I,n},\tilde{x}_{n}=x_{R,n}-ix_{I,n},p_n=p_{R,n}+ip_{I,n}, \tilde{p}_{n}=p_{R,n}-ip_{I,n}$.
Since the integrand depends only on $x_n$ and $p_n$, we can omit the integral over $\int d\tilde{x}_{n}$ and $\int d\tilde{p}_{n}$
and obtain
\begin{equation}\label{path1}
\begin{aligned}
 G\left(x_f, x_i, t\right)= \mathcal{C} \int \prod_{n=1}^{N-1} dx_{n}\prod_{n=1}^{N} dp_{n}
\exp\left\{i\sum_{n=1}^{N}\left[p_n(x_{n}-x_{n-1})-\epsilon H(p_n,x_n)\right]\right\}
 =\mathcal{C}\int_{x_i}^{x_f}\mathcal{D}[x] \mathcal{D}[p] e^{i \int_0^t\left[p \dot{x}-H\left(x, p\right)\right] d t^{\prime}},
 \end{aligned}
\end{equation}
in which the constant prefactor $\mathcal{C}$ is unimportant. The convention
of taking the limit as $N\rightarrow \infty, N\epsilon=t$,
is implicitly assumed in the path integral formula.
The path integral possesses the same form
of the conventional one except that all variables here take complex values.

In the same way and using $\langle p_n |e^{-iH(\hat{x}_\eta\left(\hat{x},\hat{p}),\hat{p}_\eta(\hat{x},\hat{p})\right)\epsilon}|x_n\rangle=e^{i \left[p_n x_n-H\left(x_\eta(x_n,p_n),p_\eta(x_n,p_n)\right)\epsilon\right]}$, the propagator $G_\eta\left(x_f, x_i, t\right)=\langle x_f |e^{iH(\hat{x}_\eta(\hat{x},\hat{p}),\hat{p}_\eta(\hat{x},\hat{p}))}|x_i\rangle$ can be expressed as
\begin{equation}
\begin{aligned}
 G_\eta\left(x_f, x_i, t\right)&=\mathcal{C} \int \prod_{n=1}^{N-1} dx_{n}\prod_{n=1}^{N} dp_{n}
\exp\left\{i\sum_{n=1}^{N}\left[p_n(x_{n}-x_{n-1})-\epsilon H(x_\eta(x_n,p_n), p_\eta(x_n,p_n))\right]\right\}\\
 &=\mathcal{C}\int_{x_i}^{x_f}\mathcal{D}[x] \mathcal{D}[p] e^{i \int_0^t\left[p \dot{x}-H\left(x_\eta(x,p), p_\eta(x,p)\right)\right] d t^{\prime}}.
 \end{aligned}
\end{equation}

To derive the semiclassical $S_\eta$ symmetry [Eq.~(4) of the main text], we interpret the restriction~\eqref{g} on the propagator
in terms of the path integral as
\begin{equation}\label{sym}
\begin{aligned}
 \int_{x_i}^{x_f} \mathcal{D}[x] \mathcal{D}[p] \exp \left\{i \int_0^t\left[p \dot{x}-H\left(x_\eta(x,p), p_\eta(x,p)\right)\right] d t^{\prime}\right\}=
 \left(\int_{x_f^*}^{x_i^*} \mathcal{D}[x] \mathcal{D}[p] \exp \left\{i \int_0^{-t^*}[p \dot{x}-H(x, p)] d t^{\prime}\right\}\right)^*.
\end{aligned}
\end{equation}
The right-hand side of Eq.~\eqref{sym} can be transformed into
\begin{equation}
\begin{aligned}
 \int_{x_f}^{x_i} \mathcal{D}\left[x^*\right] \mathcal{D}\left[p^*\right] \exp \left\{-i \int_0^{-t}\left[p^* \frac{d x^*}{d t^{\prime *}}-H^*(x, p)\right] d t^{\prime *}\right\}
=  \int_{x_f}^{x_i} \mathcal{D}[x] \mathcal{D}[p] \exp \left\{-i \int_0^{-t}\left[p \frac{d x}{d t^{\prime}}-H^*\left(x^*, p^*\right)\right] d t^{\prime}\right\},
\end{aligned}
\end{equation}
by substituting integral variables as $x\rightarrow x^*, p\rightarrow p^*, t'\rightarrow t'^*$.
Further interchange of the integral limits and a shift of time by $t$ yields
\begin{equation}
\int_{x_i}^{x_f} \mathcal{D}[x] \mathcal{D}[p] \exp \left\{i \int_{-t}^0\left[p \dot{x}-H^*\left(x^*, p^*\right)\right] d t^{\prime}\right\}
= \int_{x_i}^{x_f} \mathcal{D}[x] \mathcal{D}[p] \exp \left\{i \int_0^t\left[p \dot{x}-H^*\left(x^*, p^*\right)\right] d t^{\prime}\right\}.
\end{equation}
Then Eq.~\eqref{sym} becomes
\begin{equation}
\int_{x_i}^{x_f} \mathcal{D}[x] \mathcal{D}[p] \exp \left\{i \int_0^t\left[p \dot{x}-H\left(x_\eta(x,p), p_\eta(x,p)\right)\right] d t^{\prime}\right\}=
\int_{x_i}^{x_f} \mathcal{D}[x] \mathcal{D}[p] \exp \left\{i \int_0^t\left[p \dot{x}-H^*\left(x^*, p^*\right)\right] d t^{\prime}\right\},
\end{equation}
which is Eq.~(3) of the main text. Then the classical $S_\eta$ symmetry [Eq.~(4) of the main text]
\begin{equation}\label{seta}
   S_\eta: H(x_\eta(x,p),p_\eta(x,p))=H^*(x^*,p^*),
\end{equation}
can be extracted by comparing the integrands on both sides. In contrast to its quantum counterpart in Eq.~\eqref{etasym},
in the above equality, the complex conjugation operating on $H$ \emph{does} act on its variables $x^*, p^*$.

\section{non-Hermitian Gutzwiller trace formula}

In this section, we show in detail the derivation of the Gutzwiller trace formula in the non-Hermitian regime.
It contains three steps of saddle point approximation (SPA) applied on an equal footing to
the complex integral in Eq.~(5) of the main text, which we replicate here
\begin{equation}\label{GE}
G(E)=i\int dt \ e^{i E t}\int dx_i G(x_f=x_i,x_i,t)
=i\int dt \ e^{i E t}\int dx_i \int_{x_i}^{x_i} \mathcal{D}[x]\mathcal{D}[p] e^{i\int_0^t[p\dot{x}-H(x,p)]dt'}.
\end{equation}
First, apply SPA to the phase space path integral $\int_{x_i}^{x_i}\mathcal{D}[x]\mathcal{D}[p]$ of the Green's function,
\emph{i.e.}, expand the action $\mathcal{S}[x(t), p(t)]=\int_0^t[p\dot{x}-H(x,p)]dt'$ around its saddle point. In particular,
the condition $\delta \mathcal{S}=0$ yields the complex Hamilton's canonical equations as
\begin{equation}
\dot{x}_{cl}=\frac{\partial H}{\partial p_{cl}}, \ \ \ \dot{p}_{cl}=-\frac{\partial H}{\partial x_{cl}}.
\end{equation}
The subscript ``$cl$'' denotes the semiclassical solution, which is omitted in the main text for brevity.
The Green's function then reduces to the contributions of classical orbits $x_{cl}(t')$ plus the quantum fluctuations as
\begin{equation}
G(x_f,x_i,t)\simeq e^{i \mathcal{S}[x_{cl}(t')]}\Delta_1[x_{cl}(t')],
\end{equation}
where the quantum fluctuation term is
\begin{equation}
\begin{split}
\Delta_1&=\int_{0}^{0} \mathcal{D}[x_d] \mathcal{D}[p_d] e^{ \frac{i}{2}\delta^2 \mathcal{S}[x_d(t'),p_d(t'),x_{cl}(t')]  },\\
\delta^2 \mathcal{S}&=\int_0^t\left(2p_d\dot{x}_d-A_{cl}x^2_d-2B_{cl}x_dp_d-C_{cl}p_d^2\right) d t^{\prime},\\
A_{cl}(t^{\prime})&=\frac{\partial^2 H}{\partial x_{cl}^2}, \ \ B_{cl}(t^{\prime})=\frac{\partial^2 H}{\partial x_{cl} \partial p_{cl}}, \ \ C_{cl}(t^{\prime})=\frac{\partial^2 H}{\partial p_{cl}^2}.
\end{split}
\end{equation}
Here, $x_d=x-x_{cl}, p_d=p-p_{cl}$ measure the deviations of the coordinate and momentum from
their classical trajectories. After some tedious algebra given in the Appendix~\ref{app}, the Green's function simplifies to the following compact form
\begin{equation}\label{gscl}
    G(x_f,x_i,t)=\sqrt{\frac{-1}{2\pi i\dot{x}_{cl}(t)\dot{x}_{cl}(0)\frac{\partial t}{\partial E_{cl}}}}e^{i \mathcal{S}_{cl}(x_f,x_i,t)},
\end{equation}
where $\mathcal{S}_{cl}(x_f,x_i,t)\equiv  \mathcal{S}[x_{cl}(t)]$
and $E_{cl}\equiv H$ is the energy corresponding to the
classical orbit. It can be regarded as the analytic continuation of the results for the Hermitian systems~\cite{gutzwiller1971JMP}.

To solve the non-Hermitian energy spectrum, we focus solely on the closed orbits in Eq.~\eqref{GE} where $x_f=x_i$.
Next, we apply SPA to the second integral $G(t)=\int dx_i G(x_i,x_i,t)$, which selects periodic orbits from the closed ones.
The saddle point of the integrand is determined by
\begin{equation}
\partial_{x_i} \mathcal{S}_{cl}(x_i,x_i,t)=\partial_{a} \mathcal{S}_{cl}(a,x_i,t)|_{a=x_i}+\partial_{b} \mathcal{S}_{cl}(x_i,b,t)|_{b=x_i}=p_{cl}(t)-p_{cl}(0)=0,
\end{equation}
meaning that the initial and final momenta are identical, including both their real and imaginary components.
This is just the periodic condition of classical orbits in the complex domain.


Note that the basic periods of the orbits contributing to $G(t)$ do not necessarily have to be $t$;
they can also be multiples of basic period $t/n$ $(n=1,2,3\dots)$. Taking this
into account, $G(t)$ can be expressed as
\begin{equation}
    \begin{aligned}
        G(t)=\sum_{\gamma}\sum_{n=1}^{\infty} \oint \frac{dx_{cl}}{\dot{x}_{cl}} \sqrt{\frac{-1}{2\pi i\frac{d t}{d E_{cl}}}}e^{in\mathcal{S}_{cl}^{\gamma}(t/n)}
        =\sum_{\gamma}\sum_{n=1}^{\infty} \frac{t}{n}\sqrt{\frac{-1}{2\pi i\frac{d t}{d E_{cl}}}}e^{i\left(n\oint_{C_\gamma} p_{cl}dx_{cl}-2\pi\mu n-E_{cl}t\right)},
    \end{aligned}
\end{equation}
where $\gamma$ represents distinct categories of periodic orbits (not pertaining to different multiples of the same basic periods),
$\mathcal{S}_{cl}^{\gamma}(t/n)$ is the action accumulated over the basic period along the closed path $C_\gamma$,
and the extra phase $-2\pi n \mu$ arises from the poles of $dt/dE_{cl}$ due to $\dot{x}_{cl}=0$~\cite{rajaraman1982solitons}, where the integer-valued factor $-4\mu$ is known as the Maslov index.

Finally, we apply SPA again to calculate $G(E)=i\int dt e^{iEt}G(t)$.
Using the substitution $\tilde{t}=t/n$, $G(E)$ becomes
\begin{equation}
        G(E)=i\sum_{\gamma}\sum_{n=1}^{\infty}\int d\tilde{t} \sqrt{n} \tilde{t}\sqrt{\frac{-1}{2\pi i\frac{d \tilde{t}}{d E_{cl}}}}e^{in\left[E\tilde{t}+\oint_{C_\gamma} p_{cl}dx_{cl}-2\pi\mu-E_{cl}\tilde{t}\right]}.
\end{equation}
The saddle point of the integrand is determined by $E=-\left[\partial_{\tilde{t}} \mathcal{S}_{cl}^\gamma(\tilde{t})\right]_{\tilde{t}=T_\gamma}=E_{cl}(T_\gamma)$,
which selects periodic orbits with energy equal to the parameter $E$, and various travel times $nT_\gamma$ for different $n$ and $\gamma$. After integral over $\tilde{t}$ we finally arrive at the trace formula as
    \begin{equation}
        \begin{aligned}
        G(E)=i\sum_\gamma \sum_{n=1}^{\infty} T_{\gamma}(E)e^{-i2\pi n\mu} e^{i n \oint_{C_\gamma} p_{cl} d x_{cl}}
            =i\sum_\gamma T_{\gamma}(E)\frac{e^{i\left(\oint_{C_\gamma} p_{cl} d x_{cl}- 2\pi\mu\right)}}{1-e^{i\left(\oint_{C_\gamma} p_{cl} d x_{cl}-2\pi \mu\right)}},
        \end{aligned}
     \end{equation}
where the prefactor is partially cancelled by the contribution from the quantum fluctuations. For a single category of orbits, where the $\gamma$ label is
unnecessary, the above expression simplifies to Eq.~(6) of the main text.

\section{Symmetry of semiclassical orbits and its constraint on energy spectrum}
The $S_\eta$ symmetry of the classical Hamiltonian, expressed by Eq.~\eqref{seta},
imposes associated symmetric constraints on the dynamics and the corresponding orbits,
which will be demonstrated in this section.
Before proceeding, we specify
the relevant notations and expressions. The coordinate and momentum operators under the similarity
transformation $\eta$ are denoted as
\begin{equation}\label{eta}
\hat{x}_\eta=\eta \hat{x} \eta^{-1}, \ \ \hat{p}_\eta=\eta \hat{p} \eta^{-1},
\end{equation}
which satisfy the following eigenequations
\begin{equation}
\hat{x}_\eta|x_\eta\rangle=x_\eta|x_\eta\rangle,\ \ \hat{p}_\eta|p_\eta\rangle=p_\eta|p_\eta\rangle,
\end{equation}
with $x_\eta, p_\eta$ being the eigenvalues of the corresponding eigenstates $|x_\eta\rangle, |p_\eta\rangle$, respectively.
Both $\hat{x}_\eta=\hat{x}_\eta(\hat{x},\hat{p})$ and $\hat{p}_\eta=\hat{p}_\eta(\hat{x},\hat{p})$ can be generally expressed as functions of $\hat{x}$ and $\hat{p}$. In the path integral,
we need to evaluate the matrix elements such as $\langle p|\hat{x}_\eta|x\rangle, \langle p|\hat{p}_\eta|x\rangle$ in Eq.~\eqref{sym}, which are determined by the specific form of the functions
$\hat{x}_\eta=\hat{x}_\eta(\hat{x},\hat{p})$ and $\hat{p}_\eta=\hat{p}_\eta(\hat{x},\hat{p})$
or equivalently, the $\eta$ operation.
In particular, we have
\begin{equation}\label{xeta}
\frac{\langle p|\hat{x}_\eta(\hat{x},\hat{p})|x\rangle}{\langle p|x\rangle}={x}_\eta({x},{p}),\ \
\frac{\langle p|\hat{p}_\eta(\hat{x},\hat{p})|x\rangle}{\langle p|x\rangle}={p}_\eta({x},{p}),
\end{equation}
where $x_\eta(x,p)$ and $p_\eta(x,p)$ are c-numbers. In this way,
we establish a mapping between the classical coordinates and momenta
before and after the $\eta$ transformation as
\begin{equation}\label{etatrans}
\eta: \xi\rightarrow\xi_\eta(\xi),
\end{equation}
where we have denoted the phase space coordinates as
$\xi:=\{x, p\}^\mathsf{T}$ and $\xi_\eta(\xi):=\{x_\eta(\xi), p_\eta(\xi)\}^\mathsf{T}$ with $\mathsf{T}$ the transpose operation.
It should be noted that such a classical mapping is ultimately
defined by the $\eta$ similarity transformation in the quantum regime given by Eq.~\eqref{eta}.
For example, we consider $\eta=\mathcal{P}$ to be the spatial inversion operation and then
\begin{equation}
\hat{x}_\mathcal{P}=\mathcal{P} \hat{x} \mathcal{P}^{-1}=-\hat{x}, \ \ \hat{p}_\mathcal{P}=\mathcal{P} \hat{p} \mathcal{P}^{-1}=-\hat{p}.
\end{equation}
According to Eq.~\eqref{xeta}, the classical mapping in Eq.~\eqref{etatrans} is specified as
\begin{equation}
x_\mathcal{P}=-x,\ \ p_\mathcal{P}=-p.
\end{equation}

Next, we prove the following theorems:

\begin{theorem}
If $\xi_1(t)=\xi(t)$ is a solution of Hamilton's canonical equations along the time contour $t$,
$\xi_2(t^*)=\xi_\eta(\xi^*(\mp t))$ is also a solution, but along
the time contour $t^*$.
The sign ``$\mp$'' corresponds to whether the $\eta$ operation involves a transpose or not.
\end{theorem}

The proof relies on two key ingredients: (i) the semiclassical $\eta$ transformation~\eqref{etatrans} is canonical,
and (ii) the $S_\eta$ symmetry of the semiclassical Hamiltonian.

We first prove the canonical nature of the transformation $\eta$ in the semiclassical regime.
Starting with the commutator of position and momentum operators,
\begin{equation}
    [\hat{x},\hat{p}] = i,
\end{equation}
it transforms under the action of $\eta$ to
\begin{equation}
    [\hat{x}_\eta, \hat{p}_\eta] = \mp i,
\end{equation}
where the ``$\mp$'' corresponds to whether the $\eta$ operation involves a transpose or not.
Note that in the definition of $\eta$-PHS [Eq.~(1) in the main text], $\eta$ does not involve complex conjugation.
In the representation $|x_\eta\rangle$, the momentum $\hat{p}_\eta$ is expressed as $\mp \frac{1}{i}\partial_{x_\eta}$ and its eigenstates $|p_\eta\rangle$ are plane waves
\begin{equation}
\langle x_\eta | p_\eta \rangle = e^{\mp i p_{\eta} x_{\eta}}.
\end{equation}
Now consider the propagator $G(x_{\eta f},x_{\eta i},t)$ and express it in the phase space $\xi_\eta=\{x_\eta, p_\eta\}$ in a standard way as
\begin{equation}\label{G_eta}
G(x_{\eta f},x_{\eta i},t)=\int_{x_{\eta i}}^{x_{\eta f}} \mathcal{D}[x_{\eta}] \mathcal{D}[p_{\eta}] \exp \left\{i \int_0^t\left[\mp p_{\eta} \dot{x}_{\eta}-H\left(x(x_\eta,p_\eta), p(x_\eta,p_\eta)\right)\right] d t^{\prime}\right\}.
\end{equation}
The SPA for the path integral leads to the canonical equations
\begin{equation}
    \dot{x}_\eta=\mp\frac{\partial H\left(x(x_\eta,p_\eta), p(x_\eta,p_\eta)\right)}{\partial p_\eta}, \ \ \ \dot{p}_\eta=\pm\frac{\partial H\left(x(x_\eta,p_\eta), p(x_\eta,p_\eta)\right)}{\partial x_\eta}.
\end{equation}
To analyze the canonical transformations, it is convenient to express the equation of motion in the symplectic form as
\begin{equation}
    \frac{d \xi_\eta(t)}{d t}=\mp J \left[\nabla_{\xi_\eta} H(\xi(x_\eta,p_\eta))\right]^{\mathsf{T}},
\end{equation}
with $J=(0,1;-1,0)$. Changing the variables from $\xi_\eta$ to $\xi$ with the Jacobian matrix of the transformation $M_{\eta}(\xi)=\left(\frac{\partial x_\eta(x,p)}{\partial x},\frac{\partial x_\eta(x,p)}{\partial p};\frac{\partial p_\eta(x,p)}{\partial x},\frac{\partial p_\eta(x,p)}{\partial p}\right)$,
we obtain
\begin{equation}
    \frac{d \xi(t)}{d t}=\mp M_{\eta}^{-1}(\xi)J \left(M_{\eta}^{-1}(\xi)\right)^\mathsf{T}\left[\nabla_{\xi} H(\xi)\right]^{\mathsf{T}}.
\end{equation}
Here we have used the notations $\nabla_{\xi_\eta}:=(\partial_{x_\eta},\partial_{p_\eta})$ and $\nabla_\xi:=(\partial_x,\partial_p)$.
Comparing the equation of motion above and that obtained directly from the path integral
in the phase space $\xi$:
\begin{equation}\label{eomxi}
    \frac{d \xi(t)}{d t}=J \left[\nabla_\xi H(\xi)\right]^{\mathsf{T}},
\end{equation}
we conclude that the semiclassical counterpart of the similarity transformation $\eta$ in the
quantum regime is a (extended) canonical transformation which satisfies
\begin{equation}\label{canonical}
    M_{\eta}(\xi)J M_{\eta}^{\mathsf{T}}(\xi)=\mp J.
\end{equation}

Next, we prove the symmetric properties of the semiclassical orbits based on the canonical
nature of the $\eta$ transformation and the $S_\eta$ symmetry of the classical Hamiltonian. We start with
the complex conjugation of Eq.~\eqref{eomxi},
\begin{equation}
\frac{d \xi^*(t)}{d t^*}=J \left[\nabla_{\xi^*} H^*(\xi)\right]^{\mathsf{T}}.
\end{equation}
Changing the variables from $\xi^*$ to $\xi_\eta$ through $d\xi_\eta(\xi^*) =M_{\eta}(\xi^*){d\xi^*}$ yields
\begin{equation}
M^{-1}_{\eta}(\xi^*)\frac{d\xi_\eta(\xi^*(t))}{d t^*}=J M_\eta^{\mathsf{T}}(\xi^*)\left[\nabla_{\xi_\eta(\xi^*)}H^*(\xi)\right]^{\mathsf{T}},
\end{equation}
which further reduces to
\begin{equation}\label{motion_eq}
\begin{aligned}
   \frac{d \xi_\eta(\xi^*(t))}{d t^*}= M_{\eta}(\xi^*)J M_\eta^{\mathsf{T}}(\xi^*)\left[\nabla_{\xi_\eta(\xi^*)} H^*(\xi)\right]^{\mathsf{T}}.
\end{aligned}
\end{equation}
Applying the symmetry condition $H(\xi_\eta(\xi))=H^*(\xi^*)$ and inserting Eq.~\eqref{canonical} into Eq.~\eqref{motion_eq} gives
\begin{equation}
\begin{aligned}
    \frac{d \xi_\eta(\xi^*(t))}{d (\mp t^*)}&= J \left[\nabla_{\xi_\eta(\xi^*)} H(\xi_\eta(\xi^*))\right]^{\mathsf{T}}.
\end{aligned}
\end{equation}
where ``$\mp$'' stems from that in Eq.~\eqref{canonical}.

Making the substitution $t\rightarrow \mp t$, the above equation becomes the standard canonical equation as
\begin{equation}
\begin{aligned}
    \frac{d \xi_\eta(\xi^*(\mp t))}{d t^*}&= J \left[\nabla_{\xi_\eta(\xi^*)} H(\xi_\eta(\xi^*))\right]^{\mathsf{T}}.
\end{aligned}
\end{equation}
Therefore, we have proved \textbf{Theorem 1} that $ \xi_2(t^*)=\xi_\eta(\xi^*(\mp t))$ is also a solution of the Hamilton's
equation, but along the time contour $t^*$.

Now applying the symmetric property to closed orbits we arrive
at the following conclusion:


\begin{theorem}
The contour integrals of the orbits $O_1(t)$ and $O_2(t^*)$ satisfy
\begin{equation}\label{px}
\oint_{O_2(t^*)} p_2(t^*) dx_2(t^*)=\left(\oint_{O_1(t)} p_1(t) dx_1(t)\right)^*,
\end{equation}
where the closed orbits
$O_1(t)$ and $O_2(t^*)$ correspond to the solutions $\xi_1(t)$ and $\xi_2(t^*)$ of the canonical equation, respectively.
\end{theorem}

The proof relies on the fact that the closed orbits $O_1(t)$ and $O_2(t^*)$ form an $S_\eta$-symmetric pair. 
Applying Stokes' theorem to the left side of the above equation yields
\begin{equation}\label{o2stoks}
\oint_{O_2(t^*)} p_2(t^*) dx_2(t^*)=\int_{\Sigma_{2}} dp_2\wedge dx_2=\int_{\Sigma_{2}} dp_\eta(\xi^*)\wedge dx_\eta(\xi^*),
\end{equation}
where $\wedge$ is the wedge product, $\Sigma_{2}$ is the area encircled by $O_2(t^*)$.

The invariance of the symplectic structure, \emph{i.e.},
$M_{\eta}(\xi^*)J M_{\eta}^{\mathsf{T}}(\xi^*)=\mp J$ ensures the following Poisson brackets defined by $\xi^*$ as
\begin{equation}
    \left[x_\eta(\xi^*), p_\eta(\xi^*)\right]_{\xi^*}:= \frac{\partial x_\eta(\xi^*)}{\partial x^*}\frac{\partial p_\eta(\xi^*)}{\partial p^*}
-\frac{\partial x_\eta(\xi^*)}{\partial p^*}\frac{\partial p_\eta(\xi^*)}{\partial x^*}=\mp 1.
\end{equation}
Consequently, the wedge products transform as
\begin{equation}\label{s55}
\begin{split}
    dp_{\eta}(\xi^*)\wedge dx_{\eta}(\xi^*)&=\det\left[M_{\eta}(\xi^*)\right]dp^*\wedge dx^*
   =[x_\eta(\xi^*), p_\eta(\xi^*)]_{\xi^*}dp^*\wedge dx^*
   =\mp dp^*\wedge dx^*,
\end{split}
\end{equation}
where ``$\mp$'' again corresponds to whether $\eta$ involves a transpose or not.

The transformation from $\xi_{\eta}(\xi^*)$ to $\xi^*$ corresponds to changing the closed orbit from $O_2(t^*)$ to $O_1^*(\mp t)$.
Accordingly, the area $\Sigma_2$ encircled by $O_2(t^*)$ changes to the area $\Sigma_{1,\mp}^*$ encircled by $O_1^*(\mp t)$, where
``$\mp$'' denotes the direction of the area $\Sigma^*_1$ and the corresponding direction of the line integral along $O_1^*$.
We then have the transformation of the surface integral in phase space satisfies $\int_{\Sigma_{2}} dp_\eta(\xi^*)\wedge dx_\eta(\xi^*)=\mp\int_{\Sigma_{1,\mp}^*} dp^*\wedge dx^*$.
Combining it with Eqs.~\eqref{o2stoks} and~\eqref{s55}, we have
\begin{equation}
\begin{split}
\oint_{O_2(t^*)} p_2(t^*) dx_2(t^*)&=\mp \int_{\Sigma_{1,\mp}^*}dp^*\wedge dx^*=\mp \oint_{O_1^*(\mp t)} p^*(\mp t) dx^*(\mp t)=\left(\mp \oint_{O_1(\mp t)} p(\mp t) dx(\mp t)\right)^*\\
&=\left(\oint_{O_1(t)} p_1(t) dx_1(t)\right)^*.
\end{split}
\end{equation}
Thus, \textbf{Theorem 2} is proved.

Next, we prove the following relations between the energies and periods of symmetric orbits:
\begin{theorem}\label{th3}
 The energies and periods of two symmetric orbits, $O_1(t)$ and $O_2(t^*)$, are complex conjugates, i.e., $E_{1}=E^*_{2}$ and $T_{1}(E_1)=T^*_{2}(E_2)$.
\end{theorem}
The energies of symmetric orbits $O_1(t)$ and $O_2(t^*)$ constituting complex pairs
are indicated by the classical $S_\eta$ symmetry.
Since the Hamiltonian does not explicitly depend on time,
the energy remains constant. Therefore, we can establish the relationship between the
energies of two orbits by comparing their energies at some specific moments.
(i) For $\eta$ that does not contain a transpose, we substitute the phase space coordinates
$\xi_1(t)$ and $\xi_2(t^*)$ [corresponding to the $S_\eta$-symmetric pair $O_1(t)$ and $O_2(t^*)$] into the Hamiltonian
straightforwardly and obtain $E_1=H(\xi_1(t))=H(\xi(t))$ and $E_2=H(\xi_2(t^*))=H(\xi_\eta(\xi^*(t)))$.
Applying the $S_\eta$ symmetry of the Hamiltonian $H(\xi)=H^*(\xi_\eta(\xi^*))$ we obtain $E_1=E_2^*$.
(ii) For $\eta$ that involves a transpose, we compare the energies for $\xi_1(t)$ and $\xi_2(-t^*)$ instead.
Accordingly, we have $E_1=H(\xi_1(t))=H(\xi(t))$ and $E_2=H(\xi_2(-t^*))=H(\xi_\eta(\xi^*(t)))$, and the $S_\eta$ symmetry again leads to
$E_1=E_2^*$. Since the energy does not vary with time, the equality $E_1=E_2^*$ that holds at specific moments will hold at all times.

For the equality of periods, we consider the period of the orbit $O_1$
\begin{equation}
T_1(E_1)=\frac{\partial \oint_{O_1(t)} p_1 dx_1}{\partial E_1}.
\end{equation}
Taking the complex conjugate on both sides of the above equation and applying \textbf{Theorem 2}, we have
\begin{equation}
\begin{split}
T_1^*(E_1)&=\left(\frac{\partial \oint_{O_1(t)} p_1 dx_1}{\partial E_1}\right)^*
=\frac{\partial \left(\oint_{O_1(t)} p_1 dx_1\right)^*}{\partial E_1^*}
=\frac{\partial \oint_{O_2(t^*)} p_2 dx_2}{\partial E_2}=T_2(E_2).
\end{split}
\end{equation}
It is consistent with the fact that the time contours for the orbits $O_1(t)$
and $O_2(t^*)$ are along $t$ and $t^*$, respectively.
Thus, \textbf{Theorem 3} is proved.

By combining \textbf{Theorems 2}, \textbf{3} and the quantization condition given by the complex Gutzwiller trace formula,
we can deduce the following conclusion:
\begin{theorem}
If energy $E_n$ is an eigenenergy of the system solved by the quantization condition, then its complex conjugate $E_n^*$ is also an eigenenergy.
\end{theorem}
Specifically, we have for orbit $O_1$ that
\begin{equation}
\oint_{O_1(t)} p_1(t) dx_1(t)=(n+\mu)2\pi,
\end{equation}
which selects the orbits $O_{n1}(t)$ labeled by $n$ that satisfy the quantization condition and gives rise to discrete eigenenergies $E_{n1}=E_n$.
The symmetric counterpart of $O_{n1}(t)$ is $O_{n2}(t^*)$, whose energy is $E_{n2}=E_{n1}^*$ according to \textbf{Theorem 3}.
Moreover, \textbf{Theorem 2} ensures that the $O_{n2}(t^*)$ orbits also satisfy the quantization condition as
\begin{equation}
\oint_{O_{n2}(t^*)} p_2(t^*) dx_2(t^*)=\left(\oint_{O_{n1}(t)} p_1(t) dx_1(t)\right)^*=(n+\mu)2\pi,
\end{equation}
so that $E_{n2}=E_{n}^*$ is also an eigenenergy.

The above theorems can only yield two possible outcomes for pseudo-Hermitian eigenenergies:
\begin{theorem}
    (1) $O_1$ and $O_2$ describe the same closed orbit that possesses the $S_\eta$-symmetry and the quantized eigenenergies
    satisfying $E_n=E_n^*$, are real;
    (2) $O_1$ and $O_2$ are different orbits related by the $S_\eta$-operation ($S_\eta: O_1\leftrightarrow O_2$) and the quantized
eigenenergies constitute complex conjugate pairs with $E_{n1}=E_{n2}^*$.
\end{theorem}
\par In scenario (1), $O_1(t)$ and $O_2(t^*)$ represent the same closed orbit (denoted by $O$), meaning that they
are solutions of the same equation of motion along the same time contour, with $t=t^*$, \emph{i.e.},
along the \emph{real time} axis. Under this premise, the two solutions can still
have an arbitrary time difference $\Delta t$ such that $O_1(t)=O_2(t+\Delta t)$, corresponding
to the same closed orbit but with different initial conditions. Combining this with \textbf{Theorem 1} which states
$O_2(t) = \xi_\eta(O_1^*(\mp t))$ gives rise to the $S_\eta$-symmetry of the orbit:
\begin{equation}
    O(t-\Delta t) = \xi_\eta(O^*(\mp t)).
\end{equation}
Given that the two solutions represent the same orbit, we have for the energy that
\begin{equation}
    E_1 =H(O_1(t))=H(O_2(t+\Delta t))= E_2.
\end{equation}
Combining this with the result of \textbf{Theorem 3}, which states $E_1 = E_2^*$,
we conclude that the energy for symmetric orbits must be real.
The eigenenergies are selected from these orbits by the quantization condition
so that they must be real with $E_{n} = E_{n}^*$.

\par In scenario (2), $O_1$ and $O_2$ are different orbits that constitute a $S_\eta$-symmetric pair ($S_\eta: O_1\leftrightarrow O_2$). Then \textbf{Theorem 4}
ensures that the quantized eigenenergies constitute complex conjugate pairs with $E_{n1}=E_{n2}^*$~\textcolor{blue}{\cite{F_orbit}}.

\section{Specific examples of continuous systems}
In this section, we provide detailed derivations for the examples discussed in the main text.
Before proceeding, we outline the main procedures for solving the eigenenergies.

\begin{enumerate}
  \item Derive the  $S_\eta$ symmetry of the semiclassical Hamiltonian and the orbits.
  \item Write down the equation of motion.
  \item Search for solutions of $S_\eta$-symmetric closed orbits with real energy and time, and determine the corresponding energy interval.
  \item Impose quantization condition on symmetric orbits to obtain quantized real eigenenergies within the energy interval.
  \item Search for the $S_\eta$-asymmetric periodic orbits and their
  symmetric counterparts within complex time domain in the rest of the energy interval.
  \item Impose quantization condition on these orbits to obtain complex eigenenergies.
\end{enumerate}

\subsection{Example 1: Model with non-Hermitian skin effect}
The energy spectrum of Hamiltonian
\begin{equation}
\mathcal{H}_1=(\hat{p}+i\gamma)^2+V_0|\hat{x}|,
\end{equation}
defined within $-L/2<\text{Re}(x)<L/2$ under periodic boundary condition is derived in this section.
The Hamiltonian possesses time-reversal ($\mathcal{T}$) symmetry,
or in pseudo-Hermitian language, the $\mathsf{T}$-PHS:
\begin{equation}
    \mathsf{T} \mathcal{H}_1 \mathsf{T}^{-1}=\mathcal{H}_1^{\dagger},
\end{equation}
where $\mathsf{T}$ is the transpose operation.
In particular, $\mathsf{T}$ operates on the coordinate and momentum through
\begin{equation}
    \mathsf{T} \hat{x} \mathsf{T}^{-1}=\hat{x},\ \ \ \mathsf{T} \hat{p} \mathsf{T}^{-1}=-\hat{p}.
\end{equation}
The corresponding classical transformation of $\mathsf{T}$ is
\begin{equation}
  \mathsf{T}:  \{x,p\}\rightarrow \{x_{\mathsf{T}}=x, p_{\mathsf{T}}=-p\}.
\end{equation}
According to Eq.~\eqref{seta}, the semiclassical Hamiltonian satisfies the $S_\mathsf{T}$ symmetry, expressed as
\begin{equation}
    S_\mathsf{T}: H_1(x,-p)=H_1^*(x^*,p^*).
\end{equation}
This symmetry ensures the existence of paired solutions for closed orbits, expressed as
\begin{equation}
    O_1(t)=\{x(t),p(t)\},\ \ \ O_2(t^*)=\{x^*(-t), -p^*(-t)\},
\end{equation}
where the orbits are either identical, satisfying the $S_\mathsf{T}$ symmetry, or distinct, forming an $S_\mathsf{T}$-symmetric pair.

To find specific solutions, we write down the equation of motion
\begin{equation}
\begin{aligned}
\dot{x}(t)=2(p+i\gamma),\ \ \
\dot{p}(t)=-\text{sgn}(\text{Re}(x))V_0,
\end{aligned}
\end{equation}
where $\text{sgn}(\cdot)$ is the sign function.
We first look for the orbits with $S_\mathsf{T}$ symmetry, satisfying
\begin{equation}
x(t-\Delta t)=x^*(-t),\ \ p(t-\Delta t)=-p^*(-t).
\end{equation}
For symmetric orbits, the solution can be obtained within the domains of real time and energy according to \textbf{Theorem 3}.
It is convenient to take the phase space coordinates lying at the symmetry axes as a benchmark.
Specifically, for $t=\Delta t/2$, we have $x(-\Delta t/2)=x^*(-\Delta t/2), \ p(-\Delta t/2)=-p^*(-\Delta t/2)$.
For a given real energy $E$, the initial conditions can be chosen to be $x(\Delta t/2)=E/V_0, \ p(\Delta t/2)=-i\gamma$, when $E<V_0L/2$.
However, this initial condition can not be achieved for $E>V_0L/2$ because $\text{Re}(x)$ is restricted within $[-L/2,L/2]$.
This results in different symmetric properties of orbits for $E<V_0L/2$ and $E>V_0L/2$.

(1) For $E<V_0L/2$, the solution of the Hamilton's equations is
\begin{equation}
x(t)=\left\{
    \begin{aligned}
    &-V_0t^2+2\sqrt{E}t &\ \ \ (0<t<T_1/2)\\
    &V_0t^2-6\sqrt{E}t+\frac{8E}{V_0}&\ \ \ \ \ \ (T_1/2<t<T_1)
    \end{aligned}\right.,
\end{equation}
\begin{equation}
p(t)=\left\{
    \begin{aligned}
    &-V_0t+\sqrt{E}-i\gamma &\ \ \ (0<t<T_1/2)\\
    &V_0t-3\sqrt{E}-i\gamma &\ \ \ \ \ \ \ (T_1/2<t<T_1)
    \end{aligned}\right.,
\end{equation}
as shown in Fig.~2(a2) in the main text, which describes a particle trapped within the potential barrier,
with the period of motion being $T_1=4\sqrt{E}/V_0$.
The symmetry of the orbits is revealed by $x(t)=x^*(-t-T_1/2), \ p(t)=-p^*(-t-T_1/2)$ corresponding to $\Delta t=T_1/2$.
The solution is piecewise functions given that the potential is not smooth at $x=0$.
The quantization condition reads
\begin{equation}\label{qc1_skin}
\begin{aligned}
   \int_{-x_0}^{x_0}\left(\sqrt{E-V_0|x|}-i\gamma\right) dx+\int_{x_0}^{-x_0} \left(-\sqrt{E-V_0|x|}-i\gamma\right) dx
   =2\pi \left(n+\frac{1}{2}\right),
    \end{aligned}
\end{equation}
with $x_0=E/V_0$, which yields the real eigenvalues
\begin{equation}
E_n=\left[\frac{3V_0\pi}{4}\left(n+\frac{1}{2}\right)\right]^{2/3},
\end{equation}
as shown by the purple dots in Fig.~2(a1) of the main text.
Here, the correction $1/2$ in above quantization condition arises from the Maslov index, determined by the two zero-velocity points in one orbit, with each contributing a phase shift of $\pi/2$.

(2) When $E>V_0L/2$, it is suitable to choose the initial condition
at the boundary $x(0)=-L/2$ and the corresponding momentum is $p(0)=v_0/2-i\gamma$,
where $v_0=2\sqrt{E-V_0 L/2}$. Solving the equation of motion under this initial condition yields the trajectory
\begin{equation}
\begin{aligned}
x_1(t)&=
\left\{\begin{aligned}
&-L/2+v_0t+V_0t^2\ \ \ &(0<t<T_2/2)\\
&L/2-v_0\left(T_2-t\right)-V_0\left(T_2-t\right)^2\ \ \ &(T_2/2<t<T_2)
\end{aligned}
\right.,
\end{aligned}
\end{equation}
\begin{equation}
\begin{aligned}
p_1(t)&=
\left\{\begin{aligned}
&v_0/2+V_0t+i\gamma\ \ \ &(0<t<T_2/2)\\
&v_0/2+V_0\left(T_2-t\right)+i\gamma\ \ \ &(T_2/2<t<T_2)
\end{aligned}
\right.,
\end{aligned}
\end{equation}
which describes a particle moving rightward from $-L/2$ to $L/2$ overcoming the potential barrier
within a period $T_2=2\frac{\sqrt{E}-\sqrt{E-V_0L/2}}{V_0}$.
The trajectory $O_2(t^*)$ is related to $O_1(t)$ through the symmetric operation as
\begin{equation}
    x_2(t^*)=x_1^*(-t-T_2/2), \ p_2(t^*)=-p_1^*(-t-T_2/2),
\end{equation}
which corresponds to a left-moving solution.
By inserting the two trajectories $O_1,O_2$ into the quantization condition, we obtain the equations for the eigenenergies as
\begin{equation}\label{QC_V}
\frac{2}{3V_0}E^{3/2}-\frac{2}{3V_0}(E-V_0L/2)^{3/2}\pm i\gamma L/2=\pi n,
\end{equation}
where the phase correction is $0$ because the velocity of particle within one period does not become zero.
The presence of the imaginary term $i\gamma$ implies that the above equation should be
solved in the complex domain and the solutions are shown by the red and blue crosses in Fig.~2(a1) of the main text.
The complex eigenenergies cause the coordinates
and periods of the orbits related to the energy to become complex as well,
as shown in Fig.~2(a3) of the main text.

\subsection{Example 2: Nonreciprocal lattice subject to a magnetic field}
The energy spectrum of Hamiltonian
\begin{equation}
\mathcal{H}_2=-2[t_0\cos \hat{p}_x + i\delta_x \sin \hat{p}_x + t_0 \cos(p_y-B\hat{x})],
\end{equation}
is derived in this section. The scale of the system is set to $L=2\pi/B$ and the periodic boundary condition is adopted.
As stated in the main text, the system possesses the combined mirror-time ($\mathcal{MT}$) symmetry, or equivalently, the
 $ \mathcal{M} \mathsf{T}$-PHS:
\begin{equation}
    \mathcal{M}\mathsf{T}\mathcal{H}_2(\hat{x}-p_y/B,\hat{p}_x)(\mathcal{M}\mathsf{T})^{-1}=\mathcal{H}_2^{\dagger}(\hat{x}-p_y/B,\hat{p}_x).
\end{equation}
The $\mathcal{M}\mathsf{T}$ operator acts on the coordinates and momenta through
\begin{equation}
    \mathcal{M}\mathsf{T} \hat{x} (\mathcal{M}\mathsf{T})^{-1}=\hat{x},\ \ \mathcal{M}\mathsf{T} \hat{y} (\mathcal{M}\mathsf{T})^{-1}=-\hat{y},\ \
    (\mathcal{M}\mathsf{T}) \hat{p}_x (\mathcal{M}\mathsf{T})^{-1}=-\hat{p}_x,\ \ (\mathcal{M}\mathsf{T}) \hat{p}_y (\mathcal{M}\mathsf{T})^{-1}=\hat{p}_y,
\end{equation}
which determines the semiclassical $\mathcal{M}\mathsf{T}$ transformation as
\begin{equation}
    \mathcal{M}\mathsf{T}:  \{x,y,p_x,p_y\}\rightarrow \{x_{\mathcal{M}\mathsf{T}}=x, y_{\mathcal{M}\mathsf{T}}=-y, p_x^{\mathcal{M}\mathsf{T}}=-p_x, p_y^{\mathcal{M}\mathsf{T}}=p_y\}.
\end{equation}
The semiclassical Hamiltonian possesses the $S_{\mathcal{M}\mathsf{T}}$ symmetry as
\begin{equation}
    S_{\mathcal{M}\mathsf{T}}: H_2(x-p_y/B,-p_x)=H_2^*(x^*-p_y^*/B,p_x^*).
\end{equation}
By applying the translational operation $x \rightarrow  x + {p_y}/{B} $, it reduces to
\begin{equation}
    S_{\mathcal{M}\mathsf{T}}: H_2(x,-p_x)=H_2^*(x^*,p_x^*),
\end{equation}
which gives rise to the $S_{\mathcal{M}\mathsf{T}}$-symmetric pairs of orbits
\begin{equation}
    O_1(t)=\{x(t),p_x(t)\},\ \ \ O_2(t^*)=\{x^*(-t), -p_x^*(-t)\}.
\end{equation}

Next, we write down the equation of motion
\begin{equation}
\begin{aligned}
\dot{x}(t)=2(t_0 \sin{p_x}-i\delta_x \cos{p_x}),\ \ \
\dot{p_x}(t)=-2t_0B\sin(Bx).
\end{aligned}
\end{equation}
Similarly, we first search for real energy solutions corresponding to $S_{\mathcal{M}\mathsf{T}}$-symmetric orbits, which satisfy
\begin{equation}
x(t-\Delta t)=x^*(-t),\ \ p_x(t-\Delta t)=-p_x^*(-t).
\end{equation}
Such solutions appear near the band top and bottom, where
the particle (hole and electron, respectively) is confined within the effective potential well induced by the magnetic field,
similar to \textbf{Example 1}. There
are two turning points $x_1, x_2$ satisfying $\dot{x}=0$, with $\text{Re}(x_1)<\text{Re}(x_2)$.
We take the initial conditions to be $x(0) = x_1$ and $p_x(0)=p_0$ with $p_0$ the corresponding
momentum that satisfies $E=H(x_{1},p_0)$ for a given energy $E$.
By inserting the initial conditions into the equation of motion
and solving it numerically, we obtain the symmetric periodic orbits
that describe the particle moving between two turning points as shown in Fig.~2(b2) of the main text.
By enumerating all real energies, solving the corresponding phase space coordinates, and plugging them into the following quantization condition
\begin{equation}\label{qc_magskin}
\int_{x_1}^{x_2} p_{x,+}dx+\int_{x_2}^{x_1} p_{x,-}dx=2\pi\left(n+\frac{1}{2}\right),
\end{equation}
the eigenenergies of real values that fulfill the quantization condition are selected out; see Fig.~2(b1) of the main text.
Here, $p_{x,\pm}$ denotes the right- and left-moving paths respectively, and the correction $1/2$ in the quantization condition follows the same reason as Eq.~\eqref{qc1_skin}.
Note that the imaginary part of the integral vanishes, \emph{i.e.}, $\text{Im}\left(\oint p_x \, dx\right)=0$,
which is guaranteed by the symmetry of the orbits.

\begin{figure}[!htb]
\centering
   \includegraphics[width=0.3\textwidth]{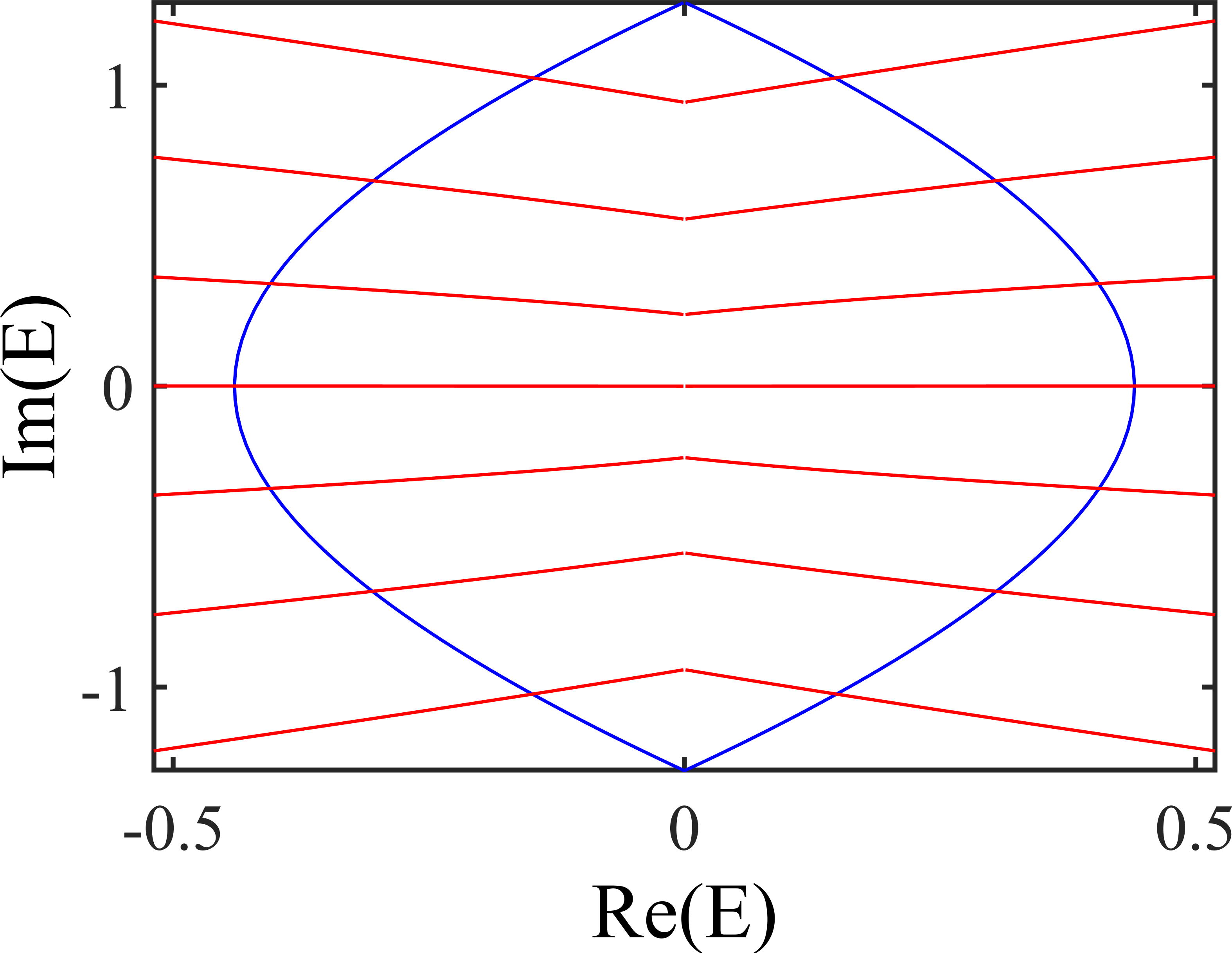}
    \caption{The curves of equation $\text{Re}\left(\int_{\mp L/2}^{\pm L/2}p_{x,\pm}dx\right)=2\pi n$(red) and $\text{Im}\left(\int_{\mp L/2}^{\pm L/2}p_{x,\pm}dx\right)=0$(blue).}
   \label{mag_solution}
\end{figure}

When the electron or hole is confined deeply in the potential well, the classical orbits chosen above
combined with the quantization condition~\eqref{qc_magskin}
yield real eigenenergies that show good consistency with the exact results obtained by directly solving the
eigenvalue equations of the Hamiltonian.
As the energy deviates from the band edges, the tunneling effect becomes more significant.
In this regime, it turns out that the periodic orbits across the potential barrier dominate the physics, as shown in Fig.~2(b3) of the main text.
The initial conditions for such kind of orbits are set to $x(0)=\pm L/2$ and $p_x(0)=p_{0,\pm}$, where $p_{0,\pm}$
is determined by the energy through $E=H(\pm L/2,p_{0,\pm})$ with ``$\pm$'' corresponding to the left- and right-moving orbits, respectively.
The time contour can be chosen as the straight line connecting $0$ and the periods $T_\pm=\pm\frac{\partial\left(\int_{-L/2}^{L/2}p_{x,\pm}dx\right)}{\partial E}$.
The orbits can be solved numerically, which are shown in Fig.~2(b3) of the main text.
The quantization conditions for the two orbits are
\begin{equation}
\begin{aligned}
\int_{\mp L/2}^{\pm L/2}p_{x,\pm}dx&=2\pi n,
\end{aligned}
\end{equation}
where the phase correction is $0$ because there are no points of zero velocity within one period.
By enumerating all complex energies and inserting the corresponding phase space coordinates
into the quantization condition, the eigenenergies can be obtained.
In particular, the quantization condition can be decomposed into
the real and imaginary parts, which read
\begin{equation}
     \text{Re}\left(\int_{-L/2}^{L/2}p_{x,\pm}dx\right)=\pm 2\pi n, \ \ \text{Im}\left(\int_{-L/2}^{L/2}p_{x,\pm}dx\right)=0.
\end{equation}
The two conditions yield two sets of curves in the complex energy plane
and the eigenenergies are determined by the intersection points of these curves, as shown in Fig.~\ref{mag_solution}.

\subsection{Example 3: Double-well potential with $\mathcal{PT}$ symmetry}\label{secvc}
The energy spectrum of Hamiltonian
\begin{equation}
\mathcal{H}_3=\hat{p}^2+g(\hat{x}^2-a^2)^2+i\Gamma \hat{x},
\end{equation}
is derived in this section.
The Hamiltonian possesses the parity-time ($\mathcal{PT}$) symmetry,
or equivalently, $\mathcal{P}\mathsf{T}$-PHS:
\begin{equation}
    \mathcal{P}\mathsf{T} \mathcal{H}_3 \mathcal{P}\mathsf{T}^{-1}=\mathcal{H}_3^{\dagger}.
\end{equation}
The $\mathcal{P}\mathsf{T}$ operation acts on the coordinate and momentum through
\begin{equation}
\mathcal{P}\mathsf{T} \hat{x} \mathcal{P}\mathsf{T}^{-1}=-\hat{x},\ \ \mathcal{P}\mathsf{T} \hat{p} \mathcal{P}\mathsf{T}^{-1}=\hat{p},
\end{equation}
which correspond to the semiclassical mapping:
\begin{equation}
    \mathcal{P}\mathsf{T}:  \{x,p\}\rightarrow \{x_{\mathcal{P}\mathsf{T}}=-x, p_{\mathcal{P}\mathsf{T}}=p\}.
\end{equation}
The semiclassical Hamilton obeys the $S_{\mathcal{P}\mathsf{T}}$ symmetry:
\begin{equation}
S_{\mathcal{P}\mathsf{T}}: H_3(-x,p)=H_3^*(x^*,p^*),
\end{equation}
which results in $S_{\mathcal{P}\mathsf{T}}$-symmetric pairs of closed
orbits
\begin{equation}
    O_1(t)=\{x(t),p(t)\},\ \ \ O_2(t^*)=\{-x^*(-t), p^*(-t)\}.
\end{equation}
The equation of motion is
\begin{equation}
\begin{aligned}
\dot{x}(t)=2p,\ \ \
\dot{p}(t)=-4gx(x^2-a^2)-i\Gamma.
\end{aligned}
\end{equation}
Again, we first solve the orbits with $S_{\mathcal{P}\mathsf{T}}$ symmetry which satisfy
\begin{equation}
x(t-\Delta t)=-x^*(-t),\ \ p(t-\Delta t)=p^*(-t).
\end{equation}
For convenience, we set the initial condition at one of the turning points.
For a given energy $E$, there are four turning points ${x^t_{1-4}}$ determined by
$E=g(x^2-a^2)^2+i\Gamma x$, which are arranged as $\text{Re}(x^t_1) \leq \text{Re}(x^t_2) \leq \text{Re}(x^t_3) \leq \text{Re}(x^t_4)$.
By inserting the initial conditions $x(0)=x^t_1,\ \ p(0)=0$ into the equation of motion and solving it numerically,
we find that the symmetry of orbits is different for
$\text{Re}(E)>V(x_0)$ and $\text{Re}(E)<V(x_0)$.
Here, the potential is $V(x)=g(x^2-a^2)^2+i\Gamma x$ and $V(x_0)$ is
the critical potential energy, corresponding to the coincidence of two turning points $x^t_2,x^t_3$.
Here, $x_0$ is defined as the solution of the equation $d V(x)/dx=0$ whose real part is the middle one among the three solutions.
The relative magnitude of $\text{Re}(E)$ and $V(x_0)$ determines whether the
particle can overcome the potential barrier at the center of the double-well potential.

(1) When $\text{Re}(E)>V(x_0)$, the orbit possesses the classical $S_{\mathcal{P}\mathsf{T}}$ symmetry,
which describes the particle moving between the two outmost turning points $x^t_1,x^t_4$,
as shown in Fig.~2(c2) of the main text.
The quantization condition reads
\begin{equation}\label{S91}
\int_{x^t_1}^{x^t_4}p_{+}dx+\int_{x^t_4}^{x^t_1}p_{-}dx=2\pi\left(n+\frac{1}{2}\right),
\end{equation}
where $p_{\pm}=\pm \sqrt{E-g(x^2-a^2)^2-i\Gamma x}$, and the phase correction is $\pi$ because there are two points of zero velocity within one period.
Zero imaginary part $\text{Im}(\oint p \, dx)=0$ is ensured by the symmetry of the orbits.
Using the same method as in \textbf{Example 2} to select the energies that satisfy
$\text{Re}(\oint p \, dx) = (n + 1/2) \pi$ in the real axis, we obtain the real eigenenergies,
as shown by the purple dots in Fig.~2(c1) of the main text.

(2) When $\text{Re}(E)<V(x_0)$, the orbit denoted as $O_1(t)$ becomes asymmetric,
which describes the particle moving between the turning points $x^t_1$ and $x^t_2$.
And its $S_{\mathcal{P}\mathsf{T}}$-symmetric counterpart $O_2(t)^*$ can be obtained through the transformation
\begin{equation}
x_2(t)=-x_1^*(-t^*),\ \ \ p_2(t)=p_1^*(-t^*),
\end{equation}
which describes the particle moving between the turning points $x^t_3$ and $x^t_4$,
as shown in Fig.2~(c3) of the main text.
Physically, the particle is deeply trapped within either the left or right potential well.
The quantization conditions for the two orbits $\oint_{O_{1,2}}p \, dx =2\pi\left(n+\frac{1}{2}\right)$ can be explicitly
expressed as
\begin{equation}
\begin{aligned}
\int_{x^t_1}^{x^t_2}p_{+}dx+\int^{x^t_1}_{x^t_2}p_{-}dx=2\pi\left(n+\frac{1}{2}\right),\ \
\int_{x^t_3}^{x^t_4}p_{+}dx+\int^{x^t_3}_{x^t_4}p_{-}dx=2\pi\left(n+\frac{1}{2}\right),
\end{aligned}\label{S93}
\end{equation}
where the phase correction is also $\pi$ for the same reason as in Eq.~\eqref{S91}.
Applying the same numerical method as in \textbf{Example 2}, the complex eigenenergies
can be obtained, indicated by red and blue crosses in Fig.~2(c1) of the main text.

\section{$\mathcal{PT}$ transition in two-level systems}\label{ptspin}

In this section, we generalize our theory to discrete non-Hermitian system,
in particular, the $\mathcal{PT}$-symmetric two-level systems, which have been
extensively explored in optic systems. We will show that the $\mathcal{PT}$
transition of the energy spectrum can be explained by the semiclassical picture as well,
similar to that for the continuous systems.
The Hamiltonian for a two-level system can be generally described by the Hamiltonian [Eq.~(10) of the main text]
\begin{equation}\label{spin}
    \mathcal{H}_4=\frac{1}{2}\bm{M}\cdot \bm{\sigma},\ \ \bm{M}=(t_1,0,i\delta_1),\ \ \bm{\sigma}=(\sigma_x,\sigma_y,\sigma_z).
    \end{equation}
The $\mathcal{PT}$ symmetry can be interpreted in terms of $\mathcal{P}\mathsf{T}$-PHS as
\begin{equation}\label{s80}
    \mathcal{P}\mathsf{T}\mathcal{H}_4(\mathcal{P}\mathsf{T})^{-1}=\mathcal{H}_4^{\dagger},
\end{equation}
where the parity operator is defined as $\mathcal{P}=\sigma_x$.

In parallel to the analysis for the continuous model, we first establish the quantum-classical correspondence
for the two-level system through the path integral formula.
Mathematically, the path integral for the two-level system is equivalent to the (pseudo-)spin path integral as~\cite{Nielsen1988NPB,altland2010condensed}
\begin{equation}\label{pi_spin}
\begin{aligned}
G(\bm{n}_f,\bm{n}_i,t)=\langle \bm{n}_f|e^{-i\mathcal{H}_4t}|\bm{n}_i\rangle
=\int_{\bm{n}_i}^{\bm{n}_f}\mathcal{D}[\bm{n}] e^{i\int_0^t \left(i\left\langle \bm{n}|
\partial_{t^{\prime}}|\bm{n}\right\rangle-\left\langle  \bm{n}| \mathcal{H}_4|\bm{n}\right\rangle\right) dt^{\prime}},
\end{aligned}
\end{equation}
which is achieved by inserting the resolution of identity in terms of the spin coherent states
$|\bm{n}\rangle=\left(\cos (\theta / 2), \sin (\theta / 2) e^{i \phi}\right)$.
Evaluating the Hamiltonian $\mathcal{H}_4$ under $|\bm{n}\rangle$ yields its classical counterpart
\begin{equation}
    H_4(\bm{n})=\langle \bm{n}|\mathcal{H}_4(\bm{\sigma})|\bm{n}\rangle=\frac{1}{2}\bm{M}\cdot \bm{n},
\end{equation}
where the unit vector $\bm{n}=(n_x, n_y, n_z)=(\sin{\theta}\cos{\phi},\sin{\theta}\sin{\phi},\cos{\theta})$ denotes the orientation of the spin.
It is noted that in the non-Hermitian scenario, an analytic continuation is performed for $\theta, \phi$ to the complex domain.
By evaluating the both sides of Eq.~\eqref{s80} in the same way we obtain the semiclassical $S_{\mathcal{P}\mathsf{T}}$ symmetry
\begin{equation}\label{spo}
   S_{\mathcal{P}\mathsf{T}}: H_4(n_x,n_y,-n_z)=H_4^*(n_x^*,n_y^*,n_z^*),
\end{equation}
where we have used
\begin{equation}
\begin{split}
    \langle \bm{n}|\mathcal{P}\mathsf{T}\mathcal{H}_4(\sigma_x,
\sigma_y,\sigma_z)(\mathcal{P}\mathsf{T})^{-1}|\bm{n}\rangle &=\langle \bm{n}|\mathcal{H}_4
(\sigma_x,\sigma_y,-\sigma_z)|\bm{n}\rangle= H_4(n_x,n_y,-n_z),\\
\langle \bm{n}|\mathcal{H}_4^{\dagger}(\bm{\sigma})|\bm{n}
\rangle&=\frac{1}{2}\langle \bm{n}|\bm{M}^*\cdot \bm{\sigma}|\bm{n}\rangle=H_4^*(n_x^*,n_y^*,n_z^*).
\end{split}
\end{equation}
This relation can also be derived through analytical
continuation. For a real $\bm{n}$, the symmetry of the semiclassical Hamiltonian
satisfies $H_4(n_x,n_y,-n_z)=\langle \bm{n}|\mathcal{P}\mathsf{T}
\mathcal{H}_4(\mathcal{P}\mathsf{T} )^{-1}|\bm{n}\rangle=\langle \bm{n}|
\mathcal{H}_4^{\dagger}|\bm{n}\rangle=H_4^*(n_x,n_y,n_z)$. Extended to the
complex domain, it reduces to Eq.~\eqref{spo}.

Next, we derive the equation of motion for this system.
By inserting the spin coherent state, the Lagrangian can be read from Eq.~\eqref{pi_spin} as
\begin{equation}\label{S_spin}
\mathcal{L}=\frac{1}{2}\left[(\cos{\theta}-1)\dot{\phi}-t_0\sin{\theta}\cos{\phi}-i\delta \cos{\theta}\right],
\end{equation}
or, in the canonical form as
\begin{equation}
\mathcal{L}=p_{\phi}\dot{\phi}-H_4(\bm{n}),\ \ p_{\phi}=\frac{\partial \mathcal{L}}{\partial \dot{\phi}}=\frac{\cos{\theta}-1}{2},
\end{equation}
where $p_{\phi}$ is the canonical momentum. The condition of stationary action yields
the semiclassical equation of motion
\begin{equation}
\left\{
\begin{aligned}
\sin{\theta} \,\dot{\theta}&=-t_1\sin{\theta}\sin{\phi},\\
\sin{\theta}\dot{\phi}&=-t_1 \cos{\theta}\cos{\phi}+i\delta_1 \sin{\theta}.
\end{aligned}\right.
\end{equation}
The physical meaning of the equation becomes clear when expressed in terms of the vector $\bm{n}$ as
\begin{equation}\label{motion_spin}
\left\{\begin{aligned}
\frac{dn_x}{dt}&=-i\delta_1 n_y,\\
\frac{dn_y}{dt}&=-(t_1n_z-i\delta_1 n_x),\\
\frac{dn_z}{dt}&=t_1 n_y,
\end{aligned}\right.
\end{equation}
using the relations $\theta=\cos^{-1}{n_z},\phi=\tan^{-1}(n_y/n_x)$.
The three equations are not independent which are restricted by $n_x^2+n_y^2+n_z^2=1$.
It can be expressed in a compact way as
\begin{equation}\label{motion_spin2}
 \frac{d\bm{n}(t)}{dt}=\bm{M}\times\bm{n}(t).
\end{equation}
The general solution $\bm{n}(t)$ of Eq.~\eqref{motion_spin2} can be written as
\begin{equation}\label{spin_orbit1}
    \left\{\begin{aligned}
        n_{x}(t)&=-\frac{i\delta_1}{2\omega}a\sin(2\omega t+\alpha)+\frac{t_1}{2\omega}c,\\
        n_{y}(t)&=a\cos(2\omega t+\alpha),\\
        n_{z}(t)&=\frac{t_1}{2\omega}a\sin(2\omega t+\alpha)+\frac{i\delta_1}{2\omega}c,
    \end{aligned}\right.
\end{equation}
where $\omega=\sqrt{t_1^2-\delta_1^2}/2$,
$a, c$ and $\alpha$ are real parameters,
and the restriction $n_x^2+n_y^2+n_z^2=1$ corresponds to $a^2+c^2=1$.
To relate the energy spectrum to the semiclassical path, we express the path integral~\eqref{pi_spin} in the coordinate $\{\phi\}$ space as
\begin{equation}
\begin{aligned}
G(\phi_f,\phi_i,t)=\langle \phi_f|e^{-i\mathcal{H}_4t}|\phi_i\rangle
=\int_{\phi_i}^{\phi_f}\mathcal{D}[\phi]\mathcal{D}[p_\phi] e^{i\int_0^t \left[p_{\phi}\dot{\phi}-H_4(\phi, p_\phi)\right] dt^{\prime}},
\end{aligned}
\end{equation}
which possesses the same formula as that of the continuous model. Then the similar trace formula can
be obtained for $G(E)$, and the energy spectrum can be solved by the quantization condition
\begin{equation}\label{qc_spin}
 \oint p_{\phi} d\phi=\oint \frac{\cos{\theta}-1}{2} d\phi=2\pi m.
\end{equation}

The $\mathcal{PT}$ transition of the energy spectrum can be interpreted by the symmetric properties of the semiclassical orbits.
We follow the same procedure and demonstrate the theorems associated with the properties
of orbits and eigenenergies in parallel to \textbf{Theorems 1-5} for the continuous models.
The $S_{\mathcal{P}\mathsf{T}}$ symmetry~\eqref{spo} of $H_4$ ensures that the solutions of the equation of motion~\eqref{motion_spin2}
fulfill the following conclusion:\\
\\
\textbf{Theorem 6.1}
\emph
{
If $\bm{n}_1(t)=\bm{n}(t)$ is a solution of equation of motion along the time contour $t$, then $\bm{n}_2(t^*):=\bm{n}_{\mathcal{P}\mathsf{T}}(\bm{n}^*(-t))=(n_x^*(-t),n_y^*(-t),-n_z^*(-t))$ is also a solution, but along the time contour $t^*$.
}\\

The proof is straightforward. Taking complex conjugation of Eq.~\eqref{motion_spin} and inverting the time yields
\begin{equation}
\left\{\begin{aligned}
\frac{dn_x^*(-t)}{dt^*}&=-i\delta_1 n_y^*,\\
\frac{dn_y^*(-t)}{dt^*}&=-(t_1(-n_z^*)-i\delta_1 n_x^*),\\
\frac{d(-n_z^*(-t))}{dt^*}&=t_1 n_y^*.
\end{aligned}\right.
\end{equation}
Comparing it with Eq.~\eqref{motion_spin},
we observe that $\bm{n}_2(t^*)=(n_x^*(-t),n_y^*(-t),-n_z^*(-t))$ is also a solution of the equation of motion, but along the time contour $t^*$,
which is the $S_{\mathcal{P}\mathsf{T}}$-symmetric counterpart of $\bm{n}_1(t)$.
According to Eq.~\eqref{spin_orbit1}, the explicit form of $\bm{n}_2(t^*)$ is given by
\begin{equation}\label{spin_orbit2}
    \left\{\begin{aligned}
        n_{2,x}(t^*)&=-\frac{i\delta_1}{2\omega^*}a\sin(2\omega^* t^*-\alpha)+\frac{t_1}{2\omega^*}c,\\
        n_{2,y}(t^*)&=a\cos(2\omega^* t^*-\alpha),\\
        n_{2,z}(t^*)&=\frac{t_1}{2\omega^*}a\sin(2\omega^* t^*-\alpha)+\frac{i\delta_1}{2\omega^*}c.
    \end{aligned}\right.
\end{equation}
\\
\textbf{Theorem 6.2}
\emph
{
The contour integrals along the orbits $\bm{n}_1(t)$ and $\bm{n}_2(t^*)$ satisfy the equality}
\begin{equation}\label{mod}
\oint_{\bm{n}_2(t^*)} \frac{\cos{\theta_2(t^*)}-1}{2} d\phi_2(t^*) = \left(\oint_{\bm{n}_1(t)} \frac{\cos{\theta_1(t)}-1}{2} d\phi_1(t)\right)^*~(\text{mod}~2\pi).
\end{equation}

This result slightly differs from Eq.~\eqref{px} for the continuous model by an additional
phase of $2\pi$ within some parametric region [see Eqs.~\eqref{realomega} and \eqref{complexomega}], which can be proved through straightforward calculation.
Specifically, we insert the expressions of $\bm{n}_1(t)$ and $\bm{n}_2(t^*)$
into the contour integral along the closed paths.
To this end, we interpret the contour integral by $\{n_x,n_y,n_z\}$ instead of $\{\theta, \phi\}$ as
\begin{equation}
\oint \frac{\cos{\theta}-1}{2} d\phi=\int_0^{T}\frac{n_z-1}{2}\frac{n_x \dot{n}_y-n_y \dot{n}_x}{n_x^2+n_y^2}dt,
\end{equation}
where $T$ is the period of spin precession.
By inserting Eq.~\eqref{spin_orbit1} and Eq.\eqref{spin_orbit2} into the above expression respectively, we obtain
\begin{equation}\label{contourintegral_spin}
\begin{aligned}
\oint_{\bm{n}_1(t)} \frac{\cos{\theta_1(t)}-1}{2} d\phi_1(t)
&= \pi\left(c-\sqrt{\frac{(2\omega+i\delta_1c)^2}{(2\omega c+i\delta_1)^2}}\frac{2\omega c+i\delta_1}{2\omega+i\delta_1c}\right), \\
\oint_{\bm{n}_2(t^*)} \frac{\cos{\theta_2(t^*)}-1}{2} d\phi_2(t^*)
&= \pi\left(c-\sqrt{\frac{(2\omega^*+i\delta_1c)^2}{(2\omega^* c+i\delta_1)^2}}\frac{2\omega^* c+i\delta_1}{2\omega^*+i\delta_1c}\right).
\end{aligned}
\end{equation}

When $\omega$ is real, Eq.~\eqref{contourintegral_spin} reduces to
\begin{equation}\label{realomega}
\begin{aligned}
\oint_{\bm{n}_1(t)} \frac{\cos{\theta_1(t)}-1}{2} d\phi_1(t)
&= \pi\left(c-\text{sgn}(c)\right), \\
\oint_{\bm{n}_2(t^*)} \frac{\cos{\theta_2(t^*)}-1}{2} d\phi_2(t^*)
&= \pi\left(c-\text{sgn}(c)\right).
\end{aligned}
\end{equation}
The discontinuity by a \(2\pi\) phase at \(c = 0\) arises from the singularity of the coherent state $|\bm{n}\rangle$ at \(\theta = \pi\).

When $\omega$ is complex, Eq.~\eqref{contourintegral_spin} can be simplified to
\begin{equation}\label{complexomega}
\begin{aligned}
\oint_{\bm{n}_1(t)} \frac{\cos{\theta_1(t)}-1}{2} d\phi_1(t)
&= \pi\left(c-\text{sgn}\left(c+\frac{2|\omega|}{\delta_1}\right)\right), \\
\oint_{\bm{n}_2(t^*)} \frac{\cos{\theta_2(t^*)}-1}{2} d\phi_2(t^*)
&= \pi\left(c-\text{sgn}\left(c-\frac{2|\omega|}{\delta_1}\right)\right).
\end{aligned}
\end{equation}
Eqs.~\eqref{realomega} and \eqref{complexomega} are unified into Eq.~\eqref{mod} of \textbf{Theorem 6.2}.
\\
\\
\textbf{Theorem 6.3}
\emph
{
 The energies and periods of two symmetric orbits, $\bm{n}_1(t)$ and $\bm{n}_2(t^*)$, are complex conjugates, i.e., $E_{1}=E^*_{2}$ and $T_{1}(E_1)=T^*_{2}(E_2)$.
}\\

By inserting the general solutions, Eq.~\eqref{spin_orbit1} and Eq.\eqref{spin_orbit2}, into the expression of energy $E=\frac{1}{2}\bm{M\cdot n}$,
one can verify that $E_1=E^*_{2}=c\omega$.
Meanwhile, the periods of two orbits satisfy $T_1(E_1)=T^*_{2}(E_2)=\pi/\omega$.\\
\\
\textbf{Theorem 6.4}
\emph
{
If energy $E_n$ is an eigenenergy of the system, then its complex conjugate $E_n^*$ is also an eigenenergy.
}\\

The relation between the contour integrals along $\bm{n}_1$ and $\bm{n}_2$ in \textbf{Theorem 6.2} combined with
the quantization condition in Eq.~\eqref{qc_spin} ensures that the quantized orbits form $S_{\mathcal{P}\mathsf{T}}$-symmetric pairs.
By applying the condition $E_1=E^*_{2}$ in \textbf{Theorem 6.3}, we prove \textbf{Theorem 6.4}.
\\
\\
\textbf{Theorem 6.5}
\emph
{
    (1) $\bm{n}_1$ and $\bm{n}_2$ describe the same closed orbit that possesses $S_{\mathcal{P}\mathsf{T}}$-symmetry and the quantized eigenenergies
    satisfying $E_n=E_n^*$, are real;
    (2) $\bm{n}_1$ and $\bm{n}_2$ are different orbits related by the $S_{\mathcal{P}\mathsf{T}}$-operation ($S_{\mathcal{P}\mathsf{T}}: \bm{n}_1\leftrightarrow \bm{n}_2$) and the quantized
eigenenergies constitute complex conjugate pairs with $E_{n1}=E_{n2}^*$.
}
\\

The proof is the same as that of \textbf{Theorem 5} for continuous models.

Next, we examine the $\mathcal{PT}$ transition through the semiclassical perspective by applying the above theorems.
From Eqs.~\eqref{spin_orbit1} and~\eqref{spin_orbit2},
it is evident that the real or complex nature of the frequency $\omega$ is closely related
to the symmetry of the orbit $\bm{n}(t)$.

(1) When $\omega$ is real, the period $T=\pi/\omega$ is also real, allowing the analysis of
semiclassical dynamics within the real time domain.
Choosing $\alpha=0$ for simplicity, we find that
$\bm{n}_2(t)=\bm{n}_{\mathcal{P}\mathsf{T}}(\bm{n}^*(-t))=\bm{n}(t)=\bm{n}_1(t)$,
indicating that $\bm{n}_2(t)$ and $\bm{n}_1(t)$ are the same orbit.
In other words, the orbit exhibits the $S_{\mathcal{P}\mathsf{T}}$ symmetry, as illustrated in Fig.~2(d2) in the main text.

(2) When $\omega$ becomes complex, the period $T=\pi/\omega$ is complex as well.
Therefore, two different orbits $\bm{n}_1(t)$ and $\bm{n}_2(t^*)$ evolve along
different time contours $t$ and $t^*$, respectively, and
form a $S_{\mathcal{P}\mathsf{T}}$-symmetric pair, as shown in Fig.~2(d3) in the main text.

The eigenenergies can be obtained by applying the quantization condition~\eqref{qc_spin} to
the semiclassical orbits. Inserting the contour integrals for $\bm{n}_1$ and $\bm{n}_2$ in Eq.~\eqref{contourintegral_spin}
into the quantization condition in Eq.~\eqref{qc_spin} yields the same result
\begin{equation}
    c=\pm 1,
\end{equation}
where $c \in [-1,1]$ is taken into account.
The corresponding eigenenergies for the two orbits are
\begin{equation}
    E_\pm=c\omega=\pm\omega,
\end{equation}
the same as the results obtained by directly diagonalizing $\mathcal{H}_4$.
Here, $\omega=\sqrt{t_1^2-\delta_1^2}/2$ is purely real or imaginary.

Note that, in the symmetry-breaking regime, the energies quantized along $\bm{n}_1$
and $\bm{n}_2$ orbits are identical.
However, this does not imply a two-fold degeneracy for each complex energy level,
because $\bm{n}_1$ with parameter $c$ and $\bm{n}_2$ with parameter $-c$ are identical.
Therefore,  double counting should be avoided to ensure that the number of
eigenvalues remains consistent with the degrees of freedom of the two-level systems. This contrasts
with continuous models, where the quantization along orbits ${O_1}$ and ${O_2}$--which are always
distinct--produces independent eigenenergies that form complex conjugate pairs in the symmetry-breaking regime.

Since only one set of orbits is independent among $\bm{n}_1$ and $\bm{n}_2$,
it is sufficient to analyze the symmetry properties of the $\bm{n}_1$ orbits only.
The semiclassical orbits selected by the quantization condition reduce to the following fixed points
\begin{equation}
    \bm{n}_{\pm}(t)=\left(\pm \frac{t_1}{2\omega},0,\pm i\frac{\delta_1}{2\omega}\right),
\end{equation}
and the real or complex nature of the eigenvalues corresponds to whether the orbits are $S_{\mathcal{P}\mathsf{T}}$-symmetric or not.
Specifically, for $\delta_1<t_1$ that corresponds to real $\omega$ and real eigenenergies,
the classical spins are along the axes $\pm \bm{M}$, possessing the $S_{\mathcal{P}\mathsf{T}}$
symmetry. When $\delta_1>t_1$, $\omega$ and the eigenenergies become purely imaginary and accordingly,
the classical spins $\bm{n}_{\pm}$ switch to the $\pm i \bm{M}$ directions. Each of them
breaks the $S_{\mathcal{P}\mathsf{T}}$ symmetry and again, they together constitute an $S_{\mathcal{P}\mathsf{T}}$-symmetric pair.

\begin{figure}[!htb]
\centering
   \includegraphics[width=0.5\textwidth]{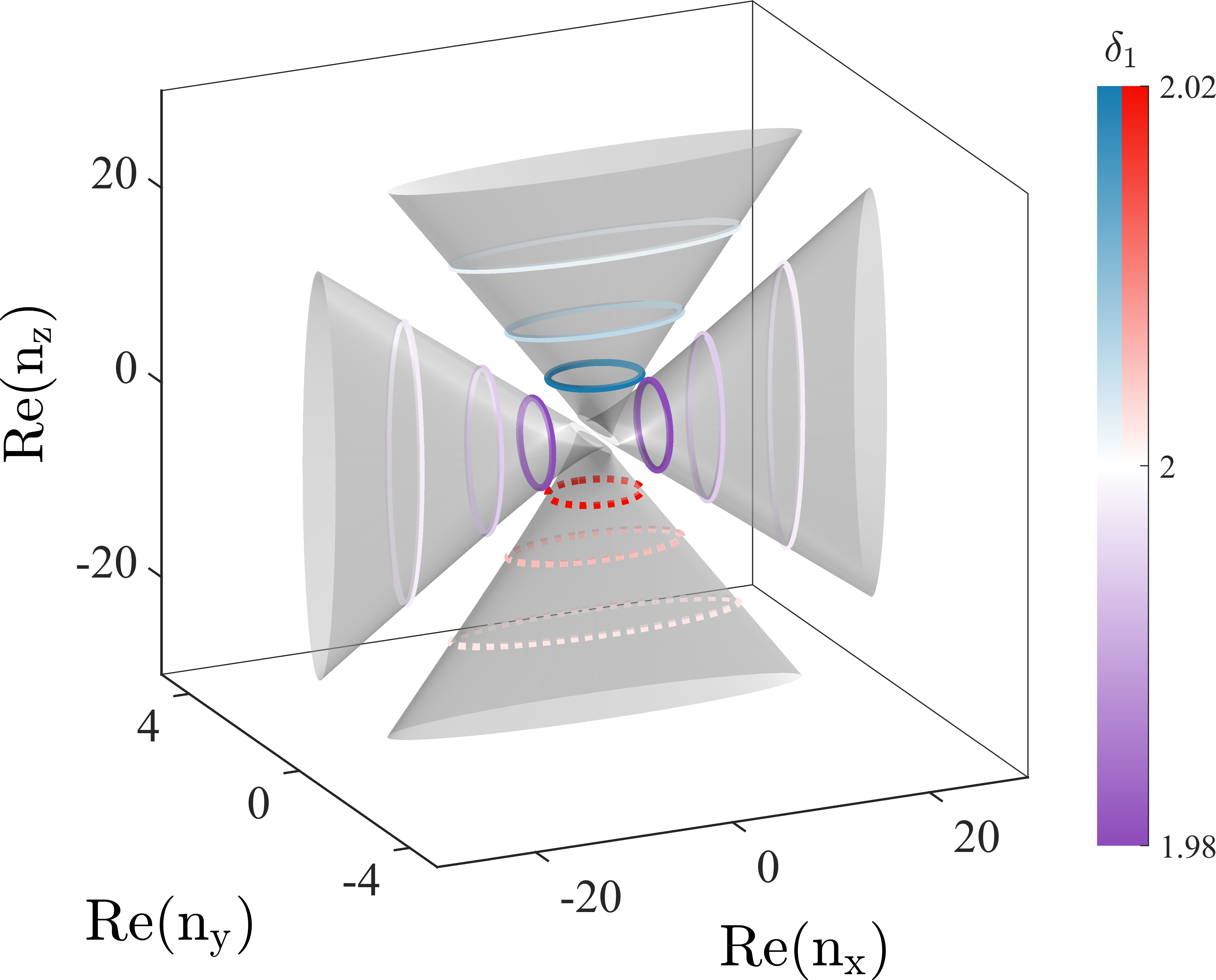}
    \caption{The semiclassical orbits of the Hamiltonian $H_4$ vary with $\delta_1$ near the critical point of $\mathcal{PT}$ transition. The corresponding $\delta_1$ values for these orbits are $1.9800, 1.9944, 1.9980, 2.0020, 2.0056$ and $2.0200$, respectively. Other orbital parameters are fixed at $t_1=2,\alpha=0$ and $c=\pm 0.8$.}
   \label{spin_orbit}
\end{figure}

The geometric features of the $\mathcal{PT}$ transition can be well illustrated by considering
general semiclassical orbits.
Specifically, as $\delta_1$ increases and approaches the critical point $t_1$, $\omega$ decreases
from a positive real value to zero, and both $n_x$
and $n_z$ diverge to infinity as shown in Fig.~\ref{spin_orbit}. Meanwhile, the average spin vector remains aligned with $\pm \bm{M}$.
As $\delta_1$ crosses $t_1$, $\omega$ transitions from real to imaginary, causing sudden changes in the real and imaginary
parts of $n_x$ and $n_z$. Accordingly, the direction of the average spin vector changes abruptly
from $\pm \bm{M}$ to $\pm i\bm{M}$.
As $\delta_1$ increases further, $n_x$ and $n_z$ decrease to
finite values, without changing the spin precession axes, as shown in Fig.~\ref{spin_orbit}.
Therefore, the divergence of the spin vector, combined with its abrupt change
in direction, signifies the $\mathcal{PT}$ transition.

\section{Physics near exceptional point}
The quantization conditions applied along semiclassical orbits yield an accurate energy spectrum in most energy regions, except near the transition point, known as the exceptional point. In fact, by carefully examining the results in Figs.~2(a1), 2(b1), and 2(c1) of the main text, small deviations between the results obtained from the quantization conditions and the exact results can be observed. These deviations become more apparent in the zoomed-in image of Fig.~2(c1), shown in Fig.~\ref{pqc}(a).
In this region, the exact energy levels converge and eventually coalesce into a single level as the system parameter is varied, as illustrated by the gray line in Fig.~\ref{E_gamma}. This behavior contrasts with that predicted by the semiclassical quantization conditions, represented by the blue dots in Fig.~\ref{E_gamma}, where adjacent energy levels maintain a finite gap.
As a result, the quantization conditions along individual periodic orbits alone are insufficient to describe the physics in the transition region, where quantum corrections must be considered. Notably, such quantum corrections can be effectively captured within the general theoretical framework developed in this work by incorporating quantum tunneling effects.

\begin{figure}[!htb]
\centering
   \includegraphics[width=0.8\textwidth]{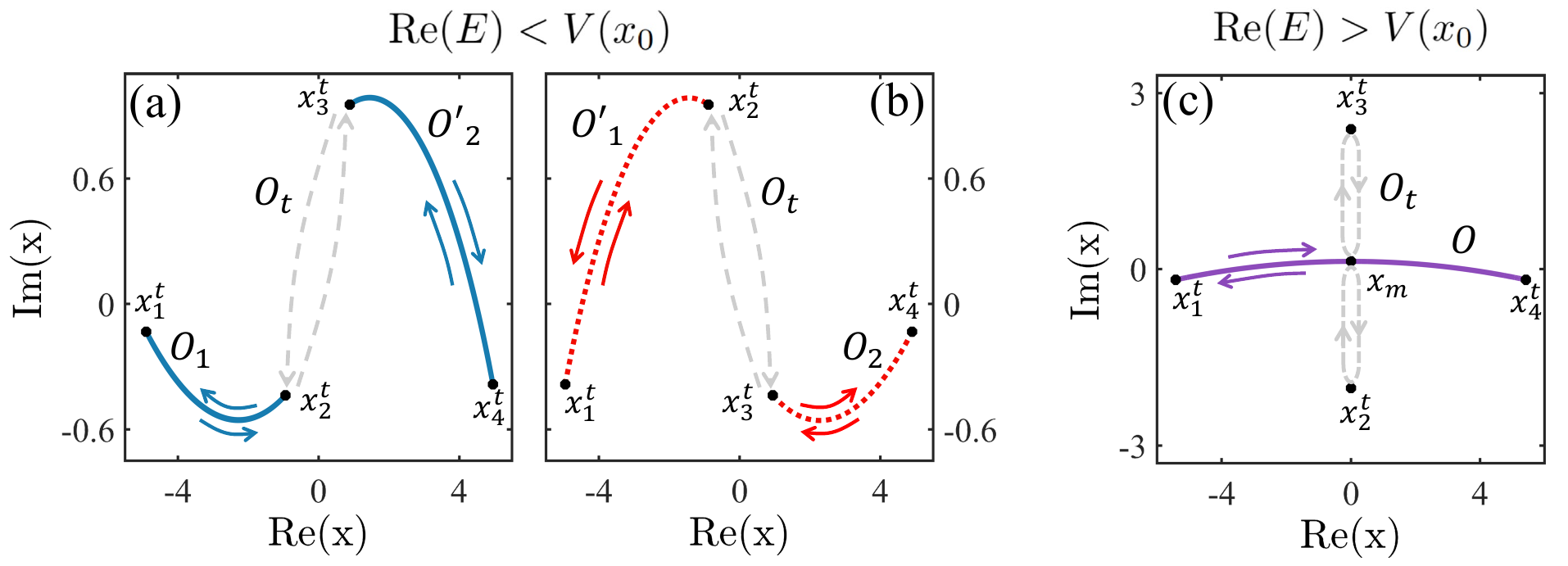}
    \caption{
        {Schematic diagram of quantum tunneling in different energy regions. The paths marked with bidirectional arrows represent classical orbits, while the gray dashed arrows indicate tunneling paths.}
    }
   \label{tunnel}
\end{figure}

\begin{figure}[!htb]
    \centering
       \includegraphics[width=0.5\textwidth]{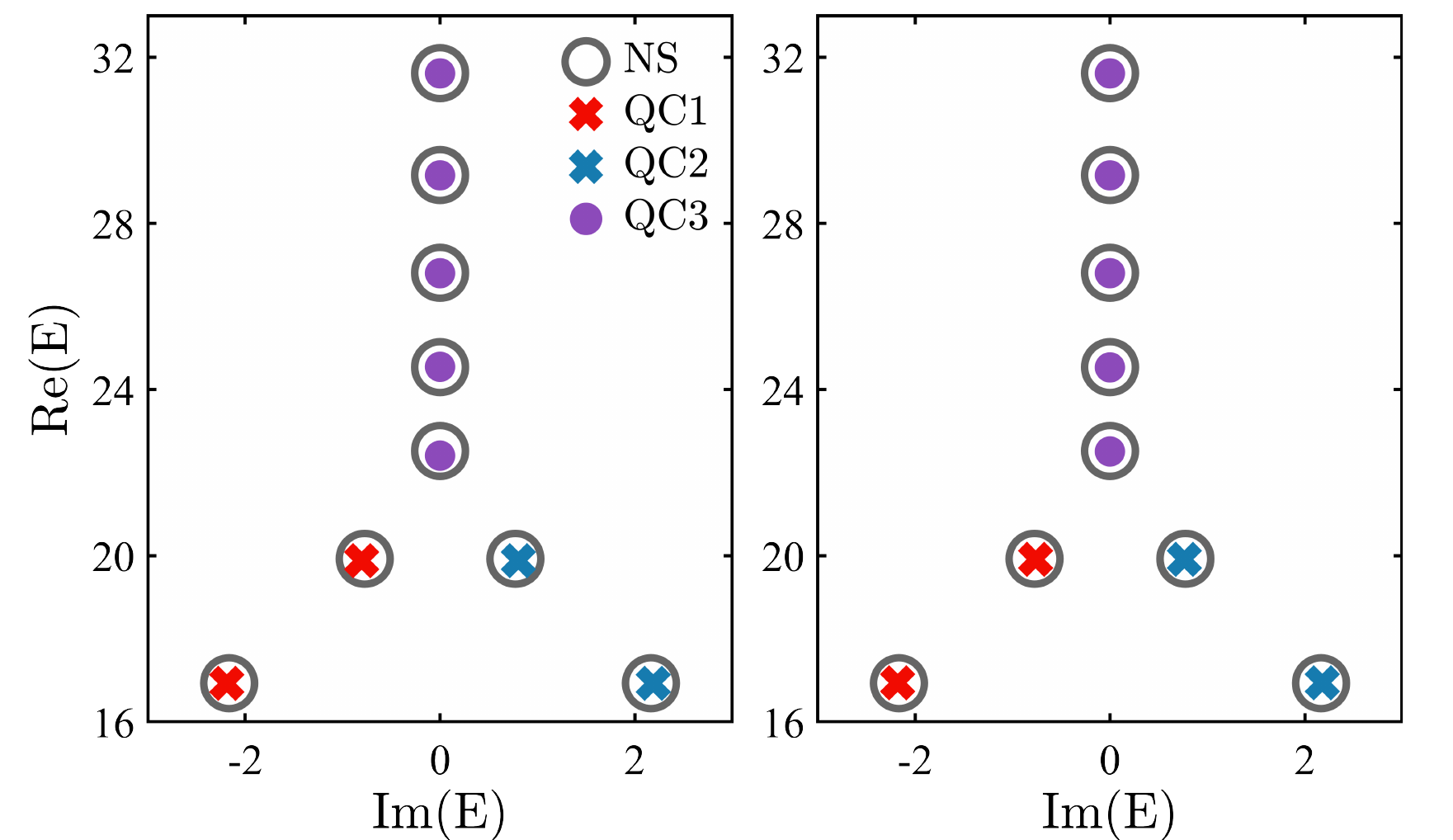}
        \caption{{Left panel: Energy levels near the transition point obtained from the quantization conditions (QC1, QC2, QC3) along individual orbits, a zoomed-in view of Fig.~2(c1) in the main text. Right panel: Energy levels solved using the generalized quantization condition~\eqref{eqc}.
        The numerical solutions (NS) from eigenvalue equation serve as a benchmark. The parameters are $a=3.5,g=0.1,\Gamma=1.25$.}
        }
       \label{pqc}
    \end{figure}

\begin{figure}[!htb]
\centering
   \includegraphics[width=0.6\textwidth]{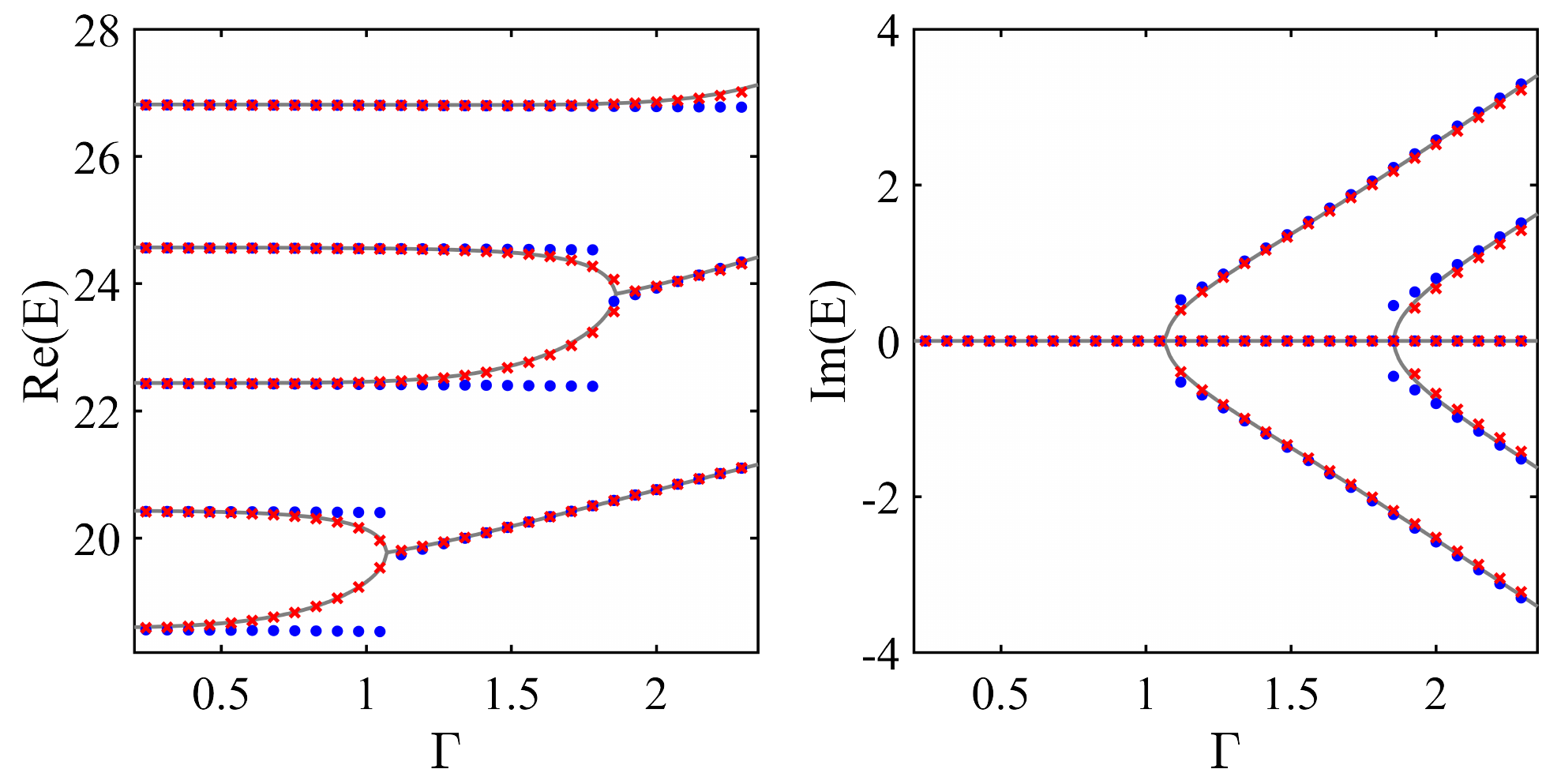}
    \caption{{The real and imaginary parts of the energy levels as functions of the parameter $\Gamma$, obtained by different methods.
    Gray lines: Numerical solutions from the eigenvalue equation.
Blue dots: Results obtained using quantization conditions along individual orbits.
Red crosses: Results from the generalized quantization condition~\eqref{eqc}.
    Other parameters are $a=3.5,g=0.1$.}
    }
   \label{E_gamma}
\end{figure}

In the following, we employ, without loss of generality, the model of double-well potential with $\mathcal{PT}$ symmetry (the third example in the main text) as an example for detailed analysis.
From the perspective of complex path integrals, the quantum corrections stem from the tunneling between classical orbits.
The earlier analysis of this model in Sec.~{\ref{secvc}} shows that semiclassical orbits are different for
$\text{Re}(E)<V(x_0)$ and $\text{Re}(E)>V(x_0)$.
Correspondingly, the tunneling behaviors are different in the two energy regions, depending on the direction from which the energy approaches $V(x_0)$.
Specifically, when $\text{Re}(E)<V(x_0)$, quantum tunneling takes place between the turning points $x^t_2$ and $x^t_3$ on two adjacent orbits: $O_1=O_1(E)$ and $O'_2=O_2(E)$ as shown in Fig.~\ref{tunnel}(a), or between the other two orbits: $O_2=O_2(E^*)$ and $O'_1=O_1(E^*)$ as shown in Fig.~\ref{tunnel}(b), each with energy conserved~\cite{O2}. 
As the energy approaches $V(x_0)$, $x^t_2$ and $x^t_3$ become closer to each other and tunneling effect becomes more pronounced.
When \( \text{Re}(E) > V(x_0) \), the classical orbits \( O_1 \), spanning from \( x^t_1 \) to \( x^t_2 \), and \( O'_2 \), extending from \( x^t_3 \) to \( x^t_4 \), merge into a single orbit \( O \) that stretches from \( x^t_1 \) to \( x^t_4 \). 
The turning points \( x^t_2 \) and \( x^t_3 \) no longer lie on the classical orbit $O$, but quantum tunneling still occurs between them.
The tunneling path \( O_t \), which extends from \( x^t_2 \) to \( x^t_3 \),  intersects with the classical orbit \( O \) at its middle point $x_m$, as illustrated in Fig.~\ref{tunnel}(c).
As the energy approaches \( V(x_0) \) from above, \( x^t_2 \) and \( x^t_3 \) become closer and the tunneling effect becomes stronger as well.

Next, we calculate the Green's function \( G(x_f, x_i, E) \) for the propagation from  \( x_i \to -\infty \) to \( x_f \to +\infty \), instead of using \( G(E) \), in order to study the quantum corrections due to tunneling effects~\cite{Carlitz1985AP}.
This approach yields the same energy spectrum as that obtained using $G(E)$, since both \( G(x_f, x_i, E) \) and \( G(E) \) share the same energy poles. Meanwhile, the information of the semiclassical orbits remains fully involved.

We begin by analyzing the regime where $\text{Re}(E)<V(x_0)$ and the tunneling between $O_1$ and $O'_2$. The tunneling between $O_2$ and $O'_1$ can be analyzed in the same way. The propagator \( G(x_f, x_i, E) \) is contributed by all Feynman paths from $x_i$ to $x_f$. These paths contain two common parts, which are direct propagations from \( x_i \) to \( x^t_1 \) and from \( x^t_4 \) to \( x_f \), along with various intermediate paths connecting \( x^t_1 \) and \( x^t_4 \). The intermediate paths contain the repeats of the basic loops of $O_1, O_2'$ and the tunneling path $O_t$ between $x^t_2$ and $x^t_3$ as shown in Fig.~\ref{tunnel}(a), in addition to the direct propagation from $x^t_1$ to $x^t_4$. 
Taking this into account and using the hierarchical procedure, $G(x_f,x_i,E)$ can be first constructed as
\begin{equation}\label{Green function}
\begin{aligned}
G(x_f,x_i,E)=e^{i\theta_f}\left\{\sum_{n=0}^{\infty}\Big[-ie^{i\theta_2}G_{1,t}({x^t_3,x^t_3})e^{i\theta_2}\Big]^n\right\} e^{i\theta_2}G_{1,t}(x^t_3,x^t_1)e^{i\theta_i}
=\frac{e^{i(\theta_i+\theta_f+\theta_2)}G_{1,t}({x^t_3,x^t_1})}{1+ie^{i2\theta_2}G_{1,t}({x^t_3,x^t_3})},
\end{aligned}
\end{equation}
where $\theta_f=\int_{x^t_4}^{x_f} p_{+} dx$ and $\theta_i=\int_{x_i}^{x^t_1} p_{+} dx$ are the phases accumulated during propagations in two terminal regions, $\theta_1=\int_{x^t_1}^{x^t_2} p_{+} dx=\int_{x^t_2}^{x^t_1} p_{-} dx$ corresponds to the one-way propagation along $O_1$, $\theta_2=\int_{x^t_3}^{x^t_4} p_{+} dx=\int_{x^t_4}^{x^t_3} p_{-} dx$ corresponds to that along $O_2'$, and the factor $-i$ arises from the phase correction due to reflection at \(x^t_4\). $G_{1,t}({x^t_3,x^t_{1(3)}})$ represents the propagator from $x^t_{1(3)}$ to $x^t_3$, contributed by all possible paths constructed by repeating $O_1$ and $O_t$, where the subscripts ``$1,t$'' indicate that the propagation occurs within the $O_1, O_t$ regions. The physical interpretation of Eq.~\eqref{Green function} is straightforward: the zeroth-order term represents the propagation from \( x^t_1 \) to \( x^t_3 \) through all paths in the $O_1$, $O_t$ regions, followed by a direct propagation from \( x^t_3 \) to \( x^t_4 \). Higher-order terms stem from repearted processes involving reflections from $x^t_4$ to $x^t_3$, propagation in the $O_1$, $O_t$ regions from $x^t_3$ to the same point, and subsequent propagation from \( x^t_3 \) to \( x^t_4 \).

Following the same spirit, the propagator $G_{1,t}({x^t_3,x^t_1})$ can be expressed as
\begin{equation}\label{G1}
    G_{1,t}({x^t_3,x^t_1}) = G_{ t}({x^t_3,x^t_2}) \left\{\sum_{n=0}^{\infty} \Big[ -ie^{i2\theta_1}G_t(x^t_2,x^t_2) \Big]^n \right\}e^{i \theta_1}
    = \frac{e^{i\theta_1}G_{t}(x^t_3,x^t_2)}{1 + i e^{i 2\theta_1} G_{t}(x^t_2,x^t_2)},
\end{equation}
where $G_{t}(x^t_{2},x^t_{2})$ and $G_{t}(x^t_{3},x^t_{2})$ represent the propagators from $x^t_2$ to $x^t_2$ and from $x^t_2$ to $x^t_3$ through all possible paths within the tunneling region by repeating $O_t$, respectively. The zeroth-order term of Eq.~\eqref{G1} represents the direct propagation from \( x^t_1 \) to \( x^t_2 \),  followed by the propagation from \( x^t_2 \) to \( x^t_3 \) through all paths in the \( O_t \) region.
Higher-order terms stem from repearted processes involving propagation in the \( O_t \) region from \( x^t_2 \) to the same point, followed by propagation along the entire $O_1$ loop.

Similarly, $G_{1,t}(x^t_3,x^t_3)$ can be expressed as

\begin{equation}\label{G2}
\begin{aligned}
G_{1,t}(x^t_3,x^t_3)&=G_{t}(x^t_3,x^t_3)+G_{1,t}({x^t_3,x^t_1})(-i)e^{i\theta_1}G_{t}(x^t_2,x^t_3)\\
&=\frac{
    G_{t}(x^t_3,x^t_3) + i e^{i2\theta_1} \left[
        G_{t}(x^t_2,x^t_2) G_{t}(x^t_3,x^t_3)
        - G_{t}(x^t_3,x^t_2)G_{t}(x^t_2,x^t_3) 
    \right]
}{
    1 + i e^{i2\theta_1} G_{t}(x^t_2,x^t_2)
},
\end{aligned}
\end{equation}
where $G_{t}(x^t_{2,3},x^t_{2,3})$ denote the propagators within the tunneling region.

Inserting Eqs.~\eqref{G1} and \eqref{G2} into Eq.~\eqref{Green function} yields the Green's function \( G(x_f, x_i, E) \) expressed in terms of all Feynman paths.
This expansion is rigorous, with the propagations along the semiclassical orbits $O_1$ and $O_2'$ described by the accumulated phases $\theta_{1,2}$, while quantum corrections are incorporated through tunneling between them.
To obtain a concrete result for the double-well potential model, we approximate the barrier $V(x)$ near $x_0$ as a quadratic potential, which yields~\cite{Carlitz1985AP}
\begin{equation}\label{G4}
\begin{aligned}
G_{t}(x^t_3,x^t_3)=G_{t}(x^t_2,x^t_2)=-i\frac{1}{\sqrt{1+e^{-2\Delta}}},\ \ \ 
G_{t}(x^t_3,x^t_2)=G_{t}(x^t_2,x^t_3)=\frac{e^{-\Delta}}{\sqrt{1+e^{-2\Delta}}},
\end{aligned}
\end{equation}
where $\Delta=-i\int_{x^t_2}^{x^t_3} p_{+} dx$. The above analysis can be applied to the tunneling between $O_2$ and $O'_1$ as shown in Fig.~\ref{tunnel}(b), by simply replacing $E$ with $E^*$ in \( G(x_f, x_i, E) \). This ensures that the energy spectrum appears as complex conjugate pairs.
Inserting Eqs.\eqref{G1}-\eqref{G4} into Eq.\eqref{Green function} yields
\begin{equation}\label{G5}
G\left(x_i, x_f, E\right) \propto \frac{e^{i(\theta_1+\theta_2)-\Delta}{(1+e^{-2 \Delta})^{-1/2}}}{1+\left(e^{i 2 \theta_1}+e^{i 2 \theta_2}\right) {(1+e^{-2 \Delta})^{-1/2}}+e^{2 i(\theta_1+\theta_2)}}.
\end{equation}

In the regime $\text{Re}(E)>V(x_0)$, the quantum correction due to tunneling in Fig.~\ref{tunnel}(c) can be analyzed in the same way. It can be shown that the propagator retains the same form as in Eq.~\eqref{G5}, provided the parameters $\theta_{1,2},\Delta$ in this energy region are defined as
    \begin{equation}\label{the12Del}
        \theta_1=\int_{x^t_1}^{x_m} p_{+} dx=\int^{x^t_1}_{x_m} p_{-} dx,\ \ \theta_2=\int_{x_m}^{x^t_4} p_{+} dx=\int^{x_m}_{x^t_4} p_{-} dx,\ \ 
        \Delta=i\int_{x^t_2}^{x^t_3} p_{+} dx,
        \end{equation}
where $x_m=(x^t_2+x^t_3)/2$.

The condition where the denominator of Eq.~\eqref{G5} equals zero defines the generalized quantization condition, valid across the entire energy range:
\begin{equation}\label{eqc}
1+\left(e^{i 2 \theta_1}+e^{i 2 \theta_2}\right) (1+e^{-2 \Delta})^{-1/2}+e^{2 i(\theta_1+\theta_2)}=0,
\end{equation}
which incorporates quantum corrections due to tunneling. 

In energy regions far from the exceptional points, the quantization condition~\eqref{eqc} reduces to those without tunneling effects. Specifically, when $\text{Re}(E)\ll V(x_0)$, we have $\Delta \rightarrow \infty$, reducing Eq.\eqref{eqc} to
\begin{equation}
\begin{aligned}
\left(1+e^{i 2 \theta_1}\right)\left(1+e^{i 2 \theta_2}\right)&=0,
\end{aligned}
\end{equation}
which coincides with the quantization condition in Eq.~\eqref{S93}.
Conversely, when $\text{Re}(E)\gg V(x_0)$, we have $\Delta \rightarrow -\infty$, reducing Eq.\eqref{eqc} to
\begin{equation}
\begin{aligned}
1+e^{2 i(\theta_1+\theta_2)}&=0,
\end{aligned}
\end{equation}
which recovers the quantization condition in Eq.~\eqref{S91} by using Eq.~\eqref{the12Del}.

For energies near the exceptional points, quantum tunneling effects lead to a finite $\Delta$, making it essential to solve Eq.~\eqref{eqc}.
By substituting the expressions of $\theta_{1,2}$ and $\Delta$, the equation transforms into one for $E=[\text{Re}(E), \text{Im}(E)]$. Its solutions yield the desired energy levels near the exceptional points. Fig.~\ref{pqc} compares the results without and with quantum corrections, shown in the left and right panels, respectively. As seen, the small deviations in energy levels near the exceptional points are effectively corrected by accounting for quantum tunneling between different orbits, demonstrating the validity of our theory. For better illustration, we plot $\text{Re}(E)$ and $\text{Im}(E)$ as functions of the model parameter $\Gamma$ in Fig.~\ref{E_gamma}, comparing three sets of results: the rigorous numerical solutions of the eigenvalue equations (gray lines), the solutions obtained using quantization conditions along individual semiclassical orbits (blue dots), and those derived from the generalized quantization condition~\eqref{eqc} (red crosses). It conveys two key insights: first, deviations in the results obtained from quantization along individual orbits occur primarily in the vicinity of the exceptional points; and second, incorporating quantum tunneling effects effectively eliminates these deviations, providing an exact description of the physics near exceptional points.

From the above analysis of the continuous model, we see that quantum tunneling plays a crucial role in the physics near exceptional points. Since quantum tunneling has no classical counterpart, this indicates that the exceptional point is inherently a quantum phenomenon without a classical interpretation.
Furthermore, in the discrete model discussed in Sec.~\ref{ptspin}, we have seen that quantum tunneling does not occur near the exceptional point (the $\mathcal{PT}$ transition point) because only a single set of orbits exists. Instead, the classical orbits diverge and change abruptly, further highlighting the breakdown of the classical description.
We thus arrive at an important conclusion: the classical hallmark of the exceptional point is its inherent lack of a classical interpretation, either due to quantum tunneling or the divergence and abrupt change of classical orbits.

\section{Appendix}\label{app}

To calculate the quantum fluctuation term $\Delta_1$, first integrating out the momentum, which yields
\begin{equation}
\begin{aligned}
\Delta_1&=\prod_{n=1}^{N}\left(\frac{1}{2\pi i \epsilon \frac{\partial \dot{x}_{cl,n}}{\partial p_{cl,n}}}\right)^{\frac{1}{2}}\int_{0}^{0} \mathcal{D}[x_d]
\exp \left\{\frac{i}{2} \int_0^t\left(a_{cl}x_d^2+2b_{cl}x_d\dot{x}_d+c_{cl}\dot{x}_d^2\right) d t^{\prime}\right\},\\
&\ \ \ \ \ \ a_{cl}(t^{\prime})=\frac{\partial^2 \mathcal{L}}{\partial x_{cl}^2}, \ \ b_{cl}(t^{\prime})=\frac{\partial^2 \mathcal{L}}{\partial x_{cl} \partial \dot{x}_{cl}}, \ \ c_{cl}(t^{\prime})=\frac{\partial^2 \mathcal{L}}{\partial \dot{x}_{cl}^2}=\frac{\partial p_{cl}}{\partial \dot{x}_{cl}},
\end{aligned}
\end{equation}
with $\mathcal{L}$ the Lagrangian. Integration by parts yields
\begin{equation}\label{3.3}
   \delta^2 \mathcal{S}=\int_0^t x_d(a_{cl}x_d-\dot{b}_{cl}x_d-\dot{c}_{cl}\dot{x}_d-c_{cl}\ddot{x}_d)  dt^{\prime}.
\end{equation}
By changing the variable as
$
y(t^{\prime})=x_d(t^{\prime})-\int_{0}^{t^{\prime}}ds \frac{\dot{f}(s)}{f(s)}x_d(s)
$~\cite{rajaraman1982solitons},
and using its inverse
\begin{equation}\label{xd}
x_d(t^{\prime})=f(t^{\prime})F(t^{\prime});\ \  F(t^{\prime})=\int_{0}^{t^{\prime}}ds \frac{\dot{y}(s)}{f(s)},
\end{equation}
with $f(t^{\prime})$ demanded to satisfy
\begin{equation}\label{f}
a_{cl}f-\dot{b}_{cl}f-\dot{c}_{cl}\dot{f}-c_{cl}\ddot{f}=0,
\end{equation}
we have
\begin{equation}\label{3.6}
    \dot{x}_d(t^{\prime})=\dot{y}(t^{\prime})+\dot{f}(t^{\prime})F(t^{\prime}),\ \
    \ddot{x}_d(t^{\prime})=\ddot{y}(t^{\prime})+\dot{f}(t^{\prime})\frac{\dot{y}(t^{\prime})}{f(t^{\prime})}+\ddot{f}(t^{\prime})F(t^{\prime}).
\end{equation}
Inserting Eqs.~\eqref{xd} and~\eqref{3.6} into Eq.~\eqref{3.3} yields
\begin{equation}\label{3.7}
   \delta^2 \mathcal{S}
   =-\int_0^{t} x_d\left(\dot{c}_{cl}\dot{y}+c_{cl}\ddot{y}+c_{cl}\dot{f}\frac{\dot{y}}{f}\right) \ dt^{\prime}
   =-\int_0^{t}\left(-c_{cl}\dot{y}\dot{f}F-c_{cl}\dot{y}f\dot{F}+c_{cl}\dot{f}\dot{y}F\right) \ dt^{\prime}
   =\int_0^t c_{cl} \dot{y}^2 dt^{\prime},
\end{equation}
which resembles a free particle in terms of the coordinate $y(t)$.
The boundary conditions are $x_d(0)=x_d(t)=0$, which become $y(0)=\ f(t)\int_{0}^{t}ds \frac{\dot{y}(s)}{f(s)}=0$
in terms of the new variable. We rewrite the path integral as
\begin{equation}
\Delta_1=\prod_{n=1}^{N}\left(\frac{1}{2\pi i \epsilon \frac{\partial \dot{x}_{cl,n}}{\partial p_{cl,n}}}\right)^{\frac{1}{2}} \int\prod_{j=1}^{N-1}dx_d^j\exp\left[\frac{i}{2}\delta^2\mathcal{S}\right],
\end{equation}
with the continuous time discretized as $t_0=0,t_1,\cdots,t_j,\cdots,t_{N-1},t_N=t$. Note that the
integrals over $x_d^0$ and $x_d^N (\equiv x_d^f)$ are absent because of the boundary conditions.
To avoid the complexity due to the explicit constraints imposed by the boundary conditions,
we incorporate its effect by inserting the following identity into $\Delta_1$,
\begin{equation}
    1=\int dx_d^f \delta(x_d^f)=\int dx_d^f \int \frac{d\alpha}{2\pi}\exp[-i\alpha x_d^f],
\end{equation}
which yields
\begin{equation}\label{3.10}
   \begin{aligned}
   \Delta_1&=\prod_{n=1}^{N}\left(\frac{1}{2\pi i \epsilon \frac{\partial \dot{x}_{cl,n}}{\partial p_{cl,n}}}\right)^{\frac{1}{2}} \int \frac{d\alpha}{2\pi}\int\prod_{j=1}^{N}dx_d^j\exp\left[-i\alpha x_d^f+\frac{i}{2}\delta^2\mathcal{S}\right]\\
   &=\prod_{n=1}^{N}\left(\frac{1}{2\pi i \epsilon \frac{\partial \dot{x}_{cl,n}}{\partial p_{cl,n}}}\right)^{\frac{1}{2}} \int \frac{d\alpha}{2\pi}\int\ |J_y|\ \prod_{j=1}^{N}dy_j  \exp\left[-i\alpha f(t)\int_{0}^{t} \frac{\dot{y}}{f} dt^{\prime} +\frac{i}{2}\int_{0}^{t} c_{cl} \dot{y}^2 \ dt^{\prime}\right]\\
   &=|J_y|\prod_{n=1}^{N}\left(\frac{1}{2\pi i \epsilon \frac{\partial \dot{x}_{cl,n}}{\partial p_{cl,n}}}\right)^{\frac{1}{2}} \int  \frac{d\alpha}{2\pi} \int\ \prod_{j=1}^{N}dy_j \exp\left[\sum_j-i\alpha f(t)\frac{y_{j}-y_{j-1}}{f_j}+i\frac{c_{cl,j}(y_{j}-y_{j-1})^2}{2\epsilon}\right],\\
\end{aligned}
\end{equation}
where $|J_y|=\sqrt{f(t)/f(0)}$ is the Jacobian determinant of the transformation $\prod_{n=1}^{N}dx_n=|J_y|\prod_{n=1}^{N}dy_n$.
Further introduce $z_j=y_{j}-y_{j-1}$, we obtain
\begin{equation}
\begin{aligned}
  \Delta_1 &=|J_y|\prod_{n=1}^{N}\left(\frac{1}{2\pi i\ \epsilon \frac{\partial \dot{x}_{cl,n}}{\partial p_{cl,n}}}\right)^{\frac{1}{2}} \int  \frac{d\alpha}{2\pi}\int\ \prod_{j=1}^{N}dz_j \exp\left[\sum_j-i\alpha f(t)\frac{z_j}{f_j}+i\frac{c_{cl,j}z_j^2}{2\epsilon}\right].
\end{aligned}
\end{equation}
The above procedures factorise the expression into a product of Gaussian integrals, which yields
\begin{equation}
\begin{aligned}
  \Delta_1 &=|J_y|\prod_{n=1}^{N}\left(\frac{1}{2\pi i \epsilon \frac{\partial \dot{x}_n}{\partial p_n}}\right)^{\frac{1}{2}} \int \frac{d\alpha}{2\pi}\prod_{j=1}^{N}\left\{-\left(\frac{2\pi i\epsilon}{c_{cl,j}}\right)^{\frac{1}{2}}\exp\left[i\frac{\alpha^2f^2(t)}{2c_{cl,j} f_j^2}\epsilon \right]\right\}\\
   &=|J_y| \int \frac{d\alpha}{2\pi}\exp\left[-i\alpha^2 f^2(t)\int_{0}^{t}\frac{1}{2c_{cl}(t^{\prime})f^2(t^{\prime})}dt^{\prime}\right]
   =\sqrt{\frac{1}{2\pi if(0)f(t)\int_{0}^{t}\frac{1}{c_{cl}(t^{\prime})f^2(t^{\prime})}dt^{\prime}}},
  \end{aligned}
\end{equation}
where equality $c_{cl}=\frac{\partial p_{cl}}{\partial \dot{x}_{cl}}$ has been used.

It is straightforward to prove $f(t^{\prime})=\dot{x}_{cl}(t^{\prime})$.
Taking the derivative of both sides of the Lagrange equation with respect to $t$, we get
\begin{equation}
a_{cl}\dot{x}_{cl}-\dot{b}_{cl}\dot{x}_{cl}-\dot{c}_{cl}\ddot{x}_{cl}-c_{cl}\dddot{x}_{cl}=0,
\end{equation}
meaning that $f$ and $\dot{x}_{cl}$ satisfy the same equation [cf.~Eq.~\eqref{f}].
So we are allowed to define $f=\dot{x}_{cl}$.
Then $\Delta_1$ reduces to
\begin{equation}
   \begin{aligned}
\Delta_1=\sqrt{\frac{1}{2\pi i\dot{x}_{cl}(t)\dot{x}_{cl}(0)\int_{0}^{t}\frac{1}{\frac{\partial p_{cl}}{\partial \dot{x}_{cl}}\dot{x}_{cl}^2}dt^{\prime}}}=\sqrt{\frac{-1}{2\pi i\dot{x}_{cl}(t)\dot{x}_{cl}(0)\frac{\partial t}{\partial E_{cl}}}},
   \end{aligned}
\end{equation}
where $E_{cl}\equiv H$ is the energy corresponding to the
classical orbit and the equalities $t
=\int_{x_i}^{x_f}\frac{\partial p_{cl}}{{\partial E_{cl}}}dx_{cl}$ and $\frac{\partial^2 p_{cl}}{{\partial E_{cl}^2}}=-{\frac{\partial^2 E_{cl}}{\partial p_{cl}^2}}/{\dot{x}_{cl}^3}$ have been employed.
The full propagator thus reduces to Eq.~\eqref{gscl}.

%
